\documentclass[aps,prx,twocolumn,amsmath,amssymb,superscriptaddress,showpacs]{revtex4-2}

\usepackage{graphicx}
\graphicspath{{Figures/}}
\usepackage{dcolumn}
\usepackage{amsmath,bm} 
\usepackage{amsbsy}
\usepackage[usenames,dvipsnames]{xcolor}
\usepackage[colorlinks=true,urlcolor=darkblue,citecolor=darkblue,linkcolor=darkred,hyperfootnotes=false]{hyperref}
\usepackage{color}
\definecolor{darkblue}{rgb}{0,0,0.6}
\definecolor{darkred}{rgb}{0.6,0,0}
\definecolor{darkgreen}{rgb}{0,0.6,0}

\usepackage{subfigure}
\usepackage[utf8]{inputenc}
\usepackage[T1]{fontenc}
\usepackage{ae,aecompl}
\usepackage{dutchcal}

\begin{document}

\title{Variational methods and deep Ritz method for active
elastic solids}

\author{Haiqin Wang}%
\affiliation{Physics Program, Guangdong Technion - Israel Institute of Technology, 241 Daxue Road, Shantou, Guangdong, China, 515063.
}%
\affiliation{Technion – Israel Institute of Technology, Haifa, Israel, 32000.
}%
\author{Boyi Zou}%
\affiliation{School of Science and Engineering, The Chinese University of Hong Kong, Shenzhen, Guangdong, China, 518172
}%
\author{Jian Su}%
\affiliation{Physics Program, Guangdong Technion - Israel Institute of Technology, 241 Daxue Road, Shantou, Guangdong, China, 515063.
}%
\author{Dong Wang}%
\affiliation{School of Science and Engineering, The Chinese University of Hong Kong, Shenzhen, Guangdong, China, 518172
}%
\affiliation{Shenzhen International Center for Industrial and Applied Mathematics, Shenzhen Research Institute of Big Data, Shenzhen, Guangdong, China, 518172
}%
\author{Xinpeng Xu} 
\email{E-mail: xu.xinpeng@gtiit.edu.cn}
\affiliation{Physics Program, Guangdong Technion - Israel Institute of Technology, 241 Daxue Road, Shantou, Guangdong, China, 515063.
}%
\affiliation{Technion – Israel Institute of Technology, Haifa, Israel, 32000.  
}%

\date{\today}

\begin{abstract}
Variational methods have been widely used in soft matter physics for both static and dynamic problems. These methods are mostly based on two variational principles: the variational principle of minimum free energy (MFEVP) and Onsager's variational principle (OVP). Our interests lie in the applications of these variational methods to active matter physics. In our former work [Soft Matter, 2021, $\bm{17}$, 3634], we have explored the applications of OVP-based variational methods for the modeling of active matter dynamics. In the present work, we explore variational (or energy) methods that are based on MFEVP for static problems in active elastic solids. We show that MFEVP can be used not only to derive equilibrium equations, but also to develop approximate solution methods, such as Ritz method, for active solid statics. Moreover, the power of Ritz-type method can be further enhanced using deep learning methods if we use deep neural networks to construct the trial functions of the variational problems. We then apply these variational methods and the deep Ritz method to study the spontaneous bending and contraction of a thin active circular plate that is induced by internal asymmetric active contraction. The circular plate is found to be bent towards its contracting side. The study of such a simple toy system gives implications for understanding the morphogenesis of solid-like confluent cell monolayers. In addition, we introduce a so-called activogravity length to characterize the importance of gravitational forces relative to internal active contraction in driving the bending of the active plate. When the lateral plate dimension is larger than the activogravity length (about 100 micron), gravitational forces become important. Such gravitaxis behaviors at multicellular scales may play significant roles in the morphogenesis and in the up-down symmetry broken during tissue development.
\end{abstract}

\maketitle


\section{Introduction}\label{Sec:Intro}

Active matter represents biological systems and their artificial analogues that are composed of large numbers of active ``agents" such as creatures and self-propelling particles moving in fluids or more complex environments~\cite{Sam2013a,Marchetti2013,Bechinger2016,Prost2015,Elgeti2015}. These constituent agents are said to be \emph{active} in the sense that they convert chemical energy continuously at the microscopic level into motion or mechanical forces. In biology, active matter includes systems at multiple scales, ranging from cell nuclei and actomyosin cytoskeletal networks at subcellular scales, to individual contractile adherent cells that are embedded in biopolymer gels, to confluent cell monolayers, to cell aggregates (or tissues) with active cell contractility, division/apoptosis, migration, and mechanical homeostasis \emph{etc}~\cite{Joanny2010,Len2013,Len2020,Julicher2017,Marchetti2019}, and even to groups of animals~\cite{Toner1995,Hemelrijk2012} and 
pedestrian crowds~\cite{Helbing2001,Castellano2009}. 
The presence of active agents in active matter breaks the detailed balance and time-reversal symmetry (TRS)~\cite{Marchetti2013,Menzel2015,Cates2015Rev,Bechinger2016}, resulting in a wealth of intriguing macroscopic structures and behaviors, such as coherent intracellular flows~\cite{Prost2015,Bechinger2016}, motility-induced phase separation~\cite{Marenduzzo2015,Cates2015Rev,Bechinger2016}, defect turbulence in
living liquid crystals, unusual mechanical and rheological properties~\cite{Sam2013a,Elgeti2015}, 
wave propagation and sustained oscillations even in the absence of inertia~\cite{Gerisch2004,Weiner2007,Inagaki2017}, flocking in animal groups \emph{etc}. 

Theoretically, there have been two major paradigms intensively explored to the study of active matter: agent-based models~\cite{Marenduzzo2015,Cates2015Rev,YangNi2020,Alt2017,Bechinger2016,Moure2021} and continuum phenomenological models~\cite{Marchetti2013,Sam2013a,Prost2015,Menzel2015,Marenduzzo2015,Cates2018,Len2013,Len2020}. The two theoretical approaches are complementary. The agent-based approach involves only a small number of parameters for each active agent, and therefore the theoretical predictions can be readily compared with experiments for some model active systems such as self-propelled colloids~\cite{ChateZhang2019}. However, the model for interacting self-propelled particles sometimes oversimplifies the problem, hence may lose some generality and applicability of its conclusions when applied to real systems, especially \emph{in vivo} biological systems~\cite{Marchetti2013,Sam2013a,Prost2015,Cates2018}. By contrast, the formulation of phenomenological models is based upon symmetry consideration, conservation laws of mass, momentum, and angular momentum, and laws of thermodynamics. 
This gives the continuum approach a much larger range of applicability and generality when applied to real biological processes~\cite{Marchetti2013,Sam2013a,Prost2015,Cates2018}. Here we focus on the continuum phenomenological models, in which active matter is represented by fields of agent density and the orientational polarization or nematic order, \emph{etc}. A continuum model for active matter is usually constructed by modifying the continuum model of a proper reference soft matter system such as micropolar fluids, liquid crystals, and gels~\cite{Marchetti2013,Sam2013a,Prost2015,Menzel2015,Marenduzzo2015,Cates2018}. In general, active matter flows at large time scales and can be modelled phenomenologically by active fluids with orientational (polar or nematic) order. However, at time scales that are smaller than the structural relaxation time (such as unbinding time of crosslinkers in cytoskeleton and characteristic time of cell migration/division/apoptosis in connective tissues), active matter is more rigid and behaves as an elastic solid~\cite{Sam2004,Sam2006,Sam2013a,Joanny2010,Ramaswamy2019}. Whereas active fluids have been studied extensively~\cite{Marchetti2013,Prost2015,QWang2020,David2021,Komura2022}, active solids have received much less attention~\cite{Sam2013a,Ramaswamy2019,Souslov2021}. 

In this work, we consider static elasticity problems in active solids. Active solids, consisting of elastically coupled active agents, combine the central properties of passive elastic solids and active fluids.
On the one hand, the positional degrees of freedom of constituent active agents have a well-defined reference state. On the other hand, activity endows these agents with an additional degree of freedom in the form of polar or nematic, active (contractile) forces. In active fluids, active agent and the environmental viscous fluids form a nonlinear hydro-active feedback loop. Active forces shear the fluids and induce flow fields, which depend on the force distribution. These flows, in turn, facilitate aligning interactions between active forces, leading to abundant collective emergent behaviors. By contrast, in active solids, active agent and the environmental elastic matrix form a nonlinear elasto-active feedback loop. Active forces deform their environmental elastic matrix and induce a strain field, which depends on the distribution of active forces. This strain field in turn reorient the agent and active forces.  
Here, we propose that variational methods based on the principle of minimum (restricted) free energy (MFEVP) provide powerful tools for the modeling and analysis of the elasto-active feedback behaviors in active solids. 
In Sec.~\ref{Sec:VarMeth}, we first introduce the general variational methods and the deep Ritz method that are based on MFEVP for static problems in active matter. Next in Sec.~\ref{Sec:ActSolid}, we use MFEVP to formulate the continuum theory of active solids and discuss the constitutive relations for both elastic and active stresses. In Sec.~\ref{Sec:App1}, we then apply the variational methods and the deep Ritz method to study the spontaneous bending and contraction of active circular plates. These studies are relevant to biological processes at multiple scales, for example, morphogenesis and gravitaxis of confluent cell monolayer, \emph{etc}. In Sec.~\ref{Sec:Conclude}, we summarize our major results and make some general remarks.


\section{Variational methods for active matter statics} \label{Sec:VarMeth}

Variational principles have been proposed in various fields of physics. Here we are particularly interested in the physics of soft and biological matter, in which two variational principles are mostly relevant: the principle of minimum free energy for static problems\cite{Doi2013,Sam2018} and Onsager's variational principle (or similar principles that extend Lagrangian variational mechanics to dissipative systems~\cite{Benjamin2021}) for dynamic problems~\cite{Xu2021,Qian2006,Doi2016,Doi2020,ChunLiu2020,QWang2020,Komura2022}. The former principle is of our particular interests in this work and is most relevant to active elastic solids. 

\subsection{The variational principle of minimum (restricted) free energy (MFEVP)} \label{Sec:VarMeth-MFEVP}

In soft and biological systems, there are usually several modes of motion with well-separated time scales, in which we can define slow and fast variables. The relaxation time of slow variables is distinctively longer than that of fast variables. In this case, the so-called restricted free energy ${\cal F}(\bm{\alpha},T)$ is introduced as an effective Hamiltonian function of a set of slow variables $\bm{\alpha}=\left(\alpha_{1}, \alpha_{2}, \ldots, \alpha_{\rm N}\right)$. In statistical mechanics, ${\cal F}(\bm{\alpha},T)$ is obtained by integrating over the fast microscopic degrees of freedom (coarse-graining) that are not far from their equilibrium state, while constraining their average to be $\bm{\alpha}$. Then the free energy is given as a functional of the probability distribution function $P(\bm{\alpha})$ as 
\begin{equation}\label{Eq:VarMeth-F}
{\cal F}(T) =\int d \bm{\alpha} \left[\beta^{-1} P(\bm{\alpha}) \ln P(\bm{\alpha})  + {\cal F}(\bm{\alpha},T)P(\bm{\alpha}) \right].
\end{equation} 
Minimization of ${\cal F}(T)$ with respect to $P(\alpha)$ yields the probability distribution function:
\begin{equation}\label{Eq:VarMeth-Peq}
P(\bm{\alpha})=Z^{-1} e^{-\beta {\cal F}(\bm{\alpha},T)}    
\end{equation}
with partition function $Z=\int d\alpha e^{-\beta {\cal F}(\bm{\alpha},T)} $, $\beta=1/k_{\rm B} T$ (here $T$ is the temperature and $k_{\rm B}$ is the Boltzmann constant), and the free energy of the system at equilibrium is given by ${\cal F}_{\rm eq}(T)=-\beta^{-1} \ln Z$. Actually, it is because of Eq.~(\ref{Eq:VarMeth-Peq}) that ${\cal F}(\bm{\alpha},T)$ is regarded as the Hamiltonian in the new coarse-grained phase space specified by $\bm{\alpha}$. We then obtain the thermodynamic variational principle of minimum (restricted) free energy (MFEVP): the most probable equilibrium state is the state $\bm{\alpha}$ which minimizes the restricted free energy ${\cal F}(\bm{\alpha},T)$, that is, ${\partial{{\cal F}}}/{\partial\bm{\alpha}}=0$. 

In active soft matter, however, the \emph{generalized thermodynamic force} $\tilde{f}_{i}$ includes not only the conservative force $\tilde{f}_{c i}(\bm{\alpha})=-{\partial {\cal F}(\bm{\alpha})}/{\partial \alpha_{i}}$ but also the \emph{active force} $\tilde{f}_{ai}(\bm{\alpha})$, which is a non-conservative force that cannot be derived from any energy function. Physically, the active forces arise from the persistent consumption of chemical energy and they continuously drive the system out of equilibrium locally at the small scale of individual active unit. For example, the active forces can be generated by biochemical reactions such as ATP hydrolysis in animal cells/tissues~\cite{Marchetti2013,Prost2015}. In this case, the total (restricted) free energy is given by
\begin{equation}\label{Eq:VarMeth-Ft}
{\cal F}_{\rm t}(\bm{\alpha})={\cal F}(\bm{\alpha})-{\cal W}(\bm{\alpha})
\end{equation}
in which ${\cal W}=\int_{\bm{\alpha}_0}^{\bm{\alpha}}\bm{\tilde{f}}(\tilde{\bm{\alpha}}) d\tilde{\bm{\alpha}}$ is the work done by ``external" forces, $\bm{\tilde{f}}$, such as active forces $\bm{\tilde{f}}_{\rm a}$ and other external forces $\bm{\tilde{f}}_{\rm ex}$, which generally depend on the state of the system specified by slow variables $\bm{\alpha}$. 
Minimization of ${\cal F}_{\rm t}$ with respect to $\bm{\alpha}$ gives the governing equilibrium (Euler-Lagrange) equations
\begin{equation}\label{Eq:VarMeth-EquilEqn}
-{\partial{{\cal F}}}/{\partial\alpha_i}+\tilde{f}_{i}=0.
\end{equation}
That is, the variational principle of minimum free energy (MFEVP) is equivalent to the balance equation of generalized forces. 

To be more specific, we consider the applications of MFEVP to elastic continuum solids where the total free energy functional is given by 
\begin{equation}\label{Eq:VarMeth-Ftot}
{\cal F}_{\rm t}[\bm{u}({\bm{r}})] = {\cal F}_{\rm e}[\bm{u}({\bm{r}})]
-\int  d\bm{r} f_{i} u_{i}-\oint dA \sigma_{{\rm s} i} u_{i}.
\end{equation}
Here $\bm{u}({\bm{r}})$ is the displacement field, ${\cal F}_{\rm e} [\bm{u}({\bm{r}})]= \int d\bm{r} F_{\rm e}(\bm{u}({\bm{r}}))$ is the deformation energy functional of the elastic solid with $F_{\rm e}(\bm{u}({\bm{r}}))$ being the energy density to be discussed in Sec.~\ref{Sec:ActSolid-Elast}. 
$\bm{f}$ and $\bm{\sigma}_{\rm s}$ are the force densities applied in the bulk and at the surfaces, respectively. 
Minimization of ${\cal F}_{\rm t}$ with respect to $\bm{u}({\bm{r}})$ gives the bulk equilibrium (or force balance) equations and boundary conditions:
\begin{subequations}\label{Eq:VarMeth-EquilEqn12}
\begin{equation}\label{Eq:VarMeth-EquilEqn1}
-{\delta{\cal F}_{\rm e}}/{\delta u_i}+f_{i}=0, \quad {\rm or} \quad \partial_j\sigma_{ij}^{\rm e}+f_i=0,
\end{equation}
\begin{equation}\label{Eq:VarMeth-EquilEqn2}
-\hat{n}_j \sigma_{ij}^{\rm e}+\sigma_{{\rm s} i}=0,\quad {\rm or} \quad u_i=u_{{\rm s}i}
\end{equation}
\end{subequations}
respectively, with $u_{{\rm s}i}$ being some given displacements at the surfaces. Here $\sigma^{\rm e}_{ij}$ is the elastic stress, generally satisfying $\delta {\cal F}_{\rm e}=\int_{\Omega} d\bm{r}\sigma^{\rm e}_{ij} \delta \epsilon_{ij}$ for both linear and nonlinear materials~\cite{Landau1986}, which will be discussed further in Sec.~\ref{Sec:ActSolid-Elast}.  
Note that the boundary conditions in Eq.~(\ref{Eq:VarMeth-EquilEqn2}) are classified into two types: \emph{essential} boundary conditions, which require the variation of $\bm{u}$ and possibly its derivatives to vanish at the boundary, and \emph{natural} boundary conditions, which require the specification of the coefficients of the variations of $\bm{u}$ and its derivatives. Natural boundary conditions can be derived directly from the minimization of energy functional, but essential boundary conditions have to be included into the energy functional as constraints, for example, by the Lagrange multiplier methods. 

\subsection{Variational methods of approximation: Ritz-type method}\label{Sec:VarMeth-Ritz}

Variational principles provide not only an equivalent substitute for the applications of governing (force balance or Euler-Lagrange) equations, but also some powerful variational methods of finding approximate solutions to these equations, \emph{e.g.}, Ritz method and the least-squares method~\cite{Reddy2017}. In these variational methods, some simple trial functions to the problem are assumed where the state variables $\bm{\alpha}=\left(\alpha_{1}, \, \alpha_{2}, \ldots \right)$ are taken as combinations of some simple functions with a much smaller number of adjustable parameters, $\bm{c}=\left(c_{1}, \, c_{2}, \ldots \right)$, \emph{i.e.},  $\bm{\alpha}=\bm{\alpha}(\bm{c})$. Then the total free energy $\mathcal{F}_{\rm t}(\bm{\alpha})$ can be written as a function of these parameters $\mathcal{F}_{\rm t}(\bm{c})$ and its minimization with respect to ${\bm{c}}$ gives approximate solutions of the static problem and determines the equilibrium state of the system. Such methods simplify the problem significantly: they bypass the derivation and solution of the complex governing Euler-Lagrange equations, and go directly from the variational statement to an approximate solution of the problem. 
These simplified solution methods are, therefore, called direct variational methods or variational methods of approximation~\cite{Reddy2017}.  

Note that in the variational method of approximation, the trial functions can be either \emph{completely empirical} arising from experiences gained from systematic numerical analysis or experimental measurements~\cite{Doi2015,Doi2020}, or assumed to \emph{linear combinations of a finite set of basis functions} such as algebraic and trigonometric polynomials~\cite{Reddy2017}. The latter choice of trial functions is known as Ritz method, in which the trial function can be approximated to arbitrary accuracy by a suitable linear combination of a sufficiently large set of basis functions. However, we would like to emphasize that no matter what forms the trial functions are assumed to be, they have to satisfy the specified essential boundary conditions (not need to satisfy the natural boundary conditions explicitly, because they are included intrinsically in the variational statement).

To be more specific and to fully demonstrate the idea of the variational methods of approximation described above, we consider the continuum elastic theory of solids~\cite{Reddy2017}, in which the state variables are only the displacement field $\bm{u}(\bm{r})$ and the total free energy is a functional of $\bm{u}(\bm{r})$, \emph{i.e.}, ${\cal F}_{\rm t}={\cal F}_{\rm t}[\bm{u}(\bm{r})]$. As mentioned above, the trial approximate solution of $\bm{u}(\bm{r})$ can be completely empirical, denoted by $\bm{U}_{\rm N}(\bm{r};\bm{c})$, which satisfies the specified essential boundary conditions and is parameterized by $N$ as yet unknown independent constant parameters $\bm{c}=(c_{1}, c_{2}, \ldots, c_{N})$. For example, we have used a one-parameter empirical (power-law) trial function in a recent work~\cite{Xu2022CPB} to explain the slow decay of cell-induced displacements measured experimentally for fibroblast spheroids in three-dimensional fibrin gels.


Besides, in the Ritz method, we seek a more explicit approximation $\bm{U}_{\rm N}(\bm{r};\bm{c})$, for a fixed and pre-selected $N$, in the finite series form of
\begin{equation}\label{Eq:VarMeth-Trial}
\bm{u}(\bm{r})\approx \bm{U}_{\rm N}(\bm{r};\bm{c}) =\sum_{i=1}^{N} c_{i} \bm{\phi}_{i}(\bm{r})+\bm{\phi}_{0}(\bm{r}),
\end{equation}
in which $\bm{\phi}_{i}(\bm{r})$ are basis functions and $c_{i}$ are the unknown independent parameters. Here in order for $\bm{U}_{\rm N}(\bm{r};\bm{c})$ to satisfy the essential boundary conditions for any $c_{i}$, we take the convenient approximation form of Eq.~(\ref{Eq:VarMeth-Trial}), requiring that $\phi_{0}(x)$ satisfies the specified essential boundary conditions of the problem, and that $\phi_{i}\, (i=1,2, \ldots, N)$ must be continuous, linearly independent, and satisfy the homogeneous form of the specified essential boundary conditions. 

In either form of the approximate trial function $\bm{U}_{\rm N}(\bm{r};\bm{c})$, substituting it into the total free energy functional ${\cal F}_{\rm t}[\bm{u}(\bm{r})]$, we obtain (after carrying out the indicated integration with respect to $\bm{r}$): ${\cal F}_{\rm t}={\cal F}_{\rm t}(\bm{c})$. Then the independent parameters $\bm{c}$ are determined by minimizing ${\cal F}_{\rm t}$ with respect to $\bm{c}$:
\begin{equation}\label{Eq:VarMeth-FtMin}
\frac{\partial {\cal F}_{\rm t}}{\partial c_{i}}=0 \quad \text { for } i=1,2, \ldots, N
\end{equation}
which represents a set of $N$ linear equations among $c_{1}, c_{2}, \ldots, c_{N}$. The solution of Eq.~(\ref{Eq:VarMeth-FtMin}) together with Eq.~(\ref{Eq:VarMeth-Trial}) or the empirical form of $\bm{U}_{\rm N}(\bm{r};\bm{c})$ yields the approximate solution $\bm{u}(\bm{r})$. This completes the description of Ritz-type variational methods of approximation.  

\subsection{Deep Ritz method (DRM): Deep learning-based methods of solving variational problems}\label{Sec:VarMeth-DeepRitz}
\begin{figure}
  \centering
  \includegraphics[width=0.9\columnwidth]{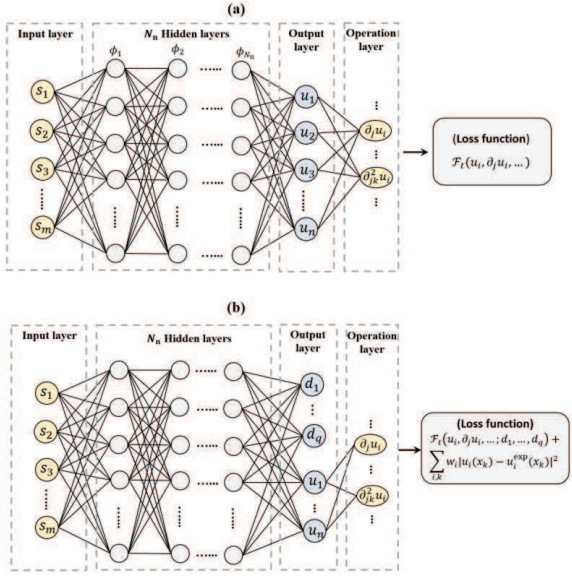} 
  \caption {Schematic illustration of the deep Ritz method as an energy minimizer that is based on deep neural networks. (a) The structure of a deep neural network with $N$ hidden layers for solving elastic problems. Each hidden layer consists of one linear transformation, activation and a skip connection as explained in Eq.~(\ref{Eq:VarMeth-MLNet}). The input of the neural network is the expanded $m$-dimension spatial coordinates $\{s_1, \ldots, s_m\}$. The output is the solution of a $n$-dimension set of slow variables $\{u_1, \ldots, u_n\}$. (b) The deep Ritz method is nested with another deep neural network to learn the material parameters $\{d_1, \ldots, d_q\}$ in the model free energy functional directly from experimental data, denoted by the slow-variable fields $u_i^{\rm exp}(x_k)$ with $x_k$ being the real spatial coordinates. Note that the weight $w_i$ in the loss function should be taken to be some large constants.
 } \label{Fig:DeepRitz}  
\end{figure}

For typical variational problems in solid elastostatics, we minimize the total free energy functional ${\cal F}_{\rm t}[\bm{u}(\bm{r})]$ with respect to the displacement field $\bm{u}(\bm{r})$. The above Ritz-type method proposes an approximate way of minimizing ${\cal F}_{\rm t}$ by taking some specific forms of admissible functions (also called trial functions) $\bm{u}(\bm{r})\approx \bm{U}(\bm{r};\bm{c})$ with some unknown adjustable parameters $\bm{c}$  (see Eq.~(\ref{Eq:VarMeth-Trial})). Using the idea of Ritz-type method, a ``deep Ritz method (DRM)''~\cite{E2018} based on deep learning has been proposed to numerically solve the above variational problems of energy-minimization. In the DRM, we use deep neural networks to construct approximate trial functions: varying the weights (or the undetermined parameters) within the network allows us to explore a rich and complex set of trial functions. The minimization of the free energy functional can then be carried out during the ``training'' phase of the network. The resulting, trained neural network is then a mapping between the spatial coordinates of the observation point (input) and the components of the state variable functions (output; here, the displacement fields). Recently, based on similar ideas, an alternative deep-learning method called ``deep energy method (DEM)'' has been proposed to solve problems of hyperelastic deformations in solid mechanics~\cite{nguyen2020deep}. 

In this work, we use DRM to solve some simple problems in active solids and employ the architecture of neural networks proposed originally by E and Yu~\cite{E2018}, which will be explained here from the following two aspects: (1) the construction of trial functions and (2) the definition and numerical treatment of the loss function (including the total free energy functional and some additional constraints or penalties). 

Firstly, the trial functions are approximated by the following composite-function form 
\begin{equation}\label{Eq:VarMeth-MLTrial}
\bm{u}(\bm{r})\approx \bm{U}_{\rm N}(\bm{r}; \bm{c})=\bm{a} \cdot [\phi_{N_{\rm n}} \circ \ldots \circ \phi_{1}(\bm{r};\tilde{c}_1)]+ \bm{b} 
\end{equation}
for a neural network with $N_{\rm n}$ hidden layers ($N_{\rm n}$ is also called the depth of the neural network, as schematically shown in Fig.~\ref{Fig:DeepRitz}(a)), in which $\phi_{i}(\bm{r};\tilde{c}_i)$ represents the $i$-th hidden layer of the neural network, typically taking the form of:
\begin{equation}\label{Eq:VarMeth-MLNet}
\phi_{i}(s;\tilde{c}_i)=\varphi(s)\left(W_{i} s+b_{i}\right)+s.
\end{equation}
Here, $\bm{a} \in \mathbb R^{n\times m}$ and $\bm{b} \in \mathbb R^{n}$ define a linear transformation that reduces the dimension (width) $m$ of the network output to the smaller dimension $n$ of the trial functions, $\bm{U}_{\rm N}(\bm{r}; \bm{c})$. In Eq.~(\ref{Eq:VarMeth-MLNet}), $\tilde{c}_i=\{W_{i} \in \mathbb{R}^{m \times m}, \, b_{i} \in \mathbb{R}^{m}\}$ denotes the parameters associated with the linear transformations in $i$-th hidden layer, $\varphi(s)$ denotes an activation function (\emph{e.g.}, ReLU, Sigmoid, and Tanh), and the last term (called residual connection, or skip connection) helps to avoid the problem of vanishing gradients, making the network much easier to be trained. 
Note that in this work the dimension $d$ of input spatial coordinates, $\bm r$, is smaller than the network dimension $m$. To resolve this discrepancy~\cite{E2018}, we pad zeros (so no undetermined parameters are introduced) to raise the input dimension from $d$ to $m$. Therefore, the full set of undetermined parameters in the trial functions is $\bm{c}=\{\tilde{c}_1,...,\tilde{c}_{N_{\rm n}},\bm{a},\bm{b}\}$ including $N=N_{\rm n}(m^2+m)+n(m+1)$ elements, as denoted in the subscript of the trial functions $\bm{U}_{\rm N}(\bm{r}; \bm{c})$.

Secondly, the loss function includes two major contributions: the total free energy functional, ${\cal F}_{\rm t}[\bm{u}(\bm{r})]$, and the terms taking into account of boundary conditions. For the functional, ${\cal F}_{\rm t}[\bm{u}(\bm{r})]$, we integrate it numerically over physical space using the Monte Carlo algorithm. More specifically, in each step of the training process, a number $N_{s}$ of spatial coordinates (the observation input points) are randomly generated from a uniform distribution and shuffled after several training steps. The energy is then calculated by summing up their values at each of the $N_{s}$ (randomly chosen) spatial coordinates. Empirically, such numerical (mesh-free) treatment for the integration of the loss function over the coordinate space is believed to be able to avoid the curse of dimensionality (\emph{i.e.}, the amount of data needed grows exponentially with the dimensionality of input coordinates), the problem of being trapped into local minimum states, and the over-fitting problems that may occur when the energy functional is discretized by any fixed spatial-grid points (for example, the Trapezoidal discretization proposed in the deep energy method, DEM~\cite{nguyen2020deep}). 
As to the contributions in the loss function from boundary conditions, since natural boundary conditions have been included intrinsically from the variational minimization of the energy functional, we only need to introduce terms (for example, by the penalty method) in the loss function to take into account of essential boundary conditions. 
Note that the spatial derivatives of trial functions in the loss function are computed by {\tt autograd} in {\tt pytorch} and the loss function is minimized by the {\tt Adam} optimizer with the stochastic gradient descent method. We refer more details to the work by Kingma and Ba~\cite{kingma2014adam}.

%

We would like to point out that the DRM is based on the combination of variational principles in physics and the deep learning method that has the capacity of mining high-dimensional information from deep neural networks. In comparison to other machine learning methods proposed particularly for active matter physics~\cite{Cichos2020,Dulaney2021,Colen2021,Zhou2021}, the DRM has several advantages as follows. (1) The DRM is naturally nonlinear, naturally adaptive, and relatively insensitive to the complexity of the energy functional. (2) In comparison to traditional numerical methods such as finite difference or finite element methods, the DRM is mesh-free and relatively insensitive to the order of the differential equation system. Moreover, the efficiency of DRM shows a much slower increase with the increasing number, $N_{\rm s}$, of spatial coordinate (input) points. In Sec.~\ref{Sec:App1-DeepRitz}, we find from our DRM studies that the training time shows linear or even sublinear increase behaviors with increasing $N_{\rm s}$. This advantage is particularly important and promising in the physics of soft matter and active matter, in which most problems involve multiple well-separated length scales and the numerical solution usually requires to cover length scales over several decades. In this case, even in a low (one to three) dimensional physical space, traditional numerical methods may become very expensive or fails completely. (3) A representation of trial functions using deep neural networks can provide a rich and complex set of trial functions, which outperforms the representation using the Ritz-type method where one has to impose carefully-chosen trial functions. (4) The DRM as an equation solver can also be nested with another optimization algorithm or simply another deep neural network to learn the material parameters in the model free energy functional directly from experimental data (as shown in Fig.~\ref{Fig:DeepRitz}(b)). Therefore, we believe that the DRM may have the potential to solve rather complex problems in active matter that involves multiple physics with multiple slow variables and multiple scales. This work of showing the validity of using DRM to solve simple problems in active solids will be the first indispensable step toward this final goal.

Before ending this subsection, we would like to emphasize that variational principles/methods should not be regarded only as an equivalent substitute for local applications of governing (equilibrium or force balance) equations. Particularly, the variational methods and the DRM presented above have many other advantages and practical importance in investigating the statics and stability of active matter as follows~\cite{Reddy2017,Landau1986,Xu2021}. 
\begin{itemize}
\item \emph{Scalar formulation}. Variational principles such as MFEVP considered here involve only physical quantities that can be defined without reference to a particular set of generalized coordinates, namely the restricted free energy and external (including active) work. This formulation is therefore automatically invariant with respect to the choice of coordinates for the system, which allows us a great flexibility in choosing state variables and in fully taking into account of symmetry requirements and constraints. 

\item \emph{Thermodynamic consistency}. The MFEVP incorporate the intrinsic structure of thermodynamics clearly. They provide compact invariant ways of obtaining thermodynamically-consistent governing equilibrium equations. Particularly in continuum field theory of elastic solids, the MFEVP provides alternate methods to the applications of local governing (equilibrium or force balance) equations in the bulk as well as matching natural boundary conditions (see Sec.~\ref{Sec:ActSolid} for details).  

\item \emph{Direct variational methods of approximation}. The direct Ritz-type variational method of finding approximation solutions for the system statics bypasses the derivation and solution of the complex Euler-Lagrange equations and goes directly from a variational statement to the solution of the problem. This approximation method helps to pick up the most important static behaviors and to simplify the calculations significantly from complicated partial differential equation systems to simple ordinary differential equations.

\item \emph{Fits well with deep-learning methods}. The Ritz-type variational problems for approximating system statics can be solved numerically using deep-learning methods. Such methods are naturally adaptive, fit well with fast stochastic gradient descent algorithms, and are relatively insensitive to the complexity of the energy functional. Particularly, mesh-free deep learning methods are known to be relatively insensitive to the order of the differential equation system and be able to avoid the problem of being trapped into local minimum states and the over-fitting problems. Moreover, their computational efficiency shows much slower increase with increasing number of spatial coordinate (input) points, which is highly promising in solving multiscale problems in soft matter and active matter physics. 
\end{itemize} 
The variational methods mentioned above have been applied successfully to many problems of elastic solids that are drawn from the problems of bars, beams, torsion, and membranes~\cite{Reddy2017,Landau1986}. In this work, we will show that these variational methods and the DRM can also be used to study the static problems in active elastic solids that are mostly motivated by cell and tissue mechanobiology~\cite{Sam2013a}.  






\section{Active elastic solids: continuum theory}\label{Sec:ActSolid}

In this section, we consider active elastic solids consisting of elastically coupled active agents that apply active forces (regarded non-conservative external forces herein). We show how to apply the above variational methods developed in the continuum theory of elasticity to active solids. 

\subsection{Total free energy and equilibrium equations}\label{Sec:ActSolid-Energy}

In active elastic solids, locally applied active forces deform their environmental elastic matrix and induce a strain field, which depends on the distribution of active forces. In general, active agents and the environmental elastic matrix form a nonlinear elasto-active feedback loop, that is, the strain field induced by active agent forces will in turn reorient the agent and active forces. However, in this work, we neglect for simplicity the backward effects of strain field on the change of active forces and only consider local active forces with a fixed and given distribution. In this case, the total free energy functional is, similarly as Eq.~(\ref{Eq:VarMeth-Ftot}), given by
\begin{equation}\label{Eq:ActSolid-Ftot}
{\cal F}_{\rm t}[\bm{u}({\bm{r}})] = {\cal F}_{\rm e}[\bm{u}({\bm{r}})]-
\sum_{n=1}^{N} \int_{\Omega_n} d{\bm r}f_{{\rm a}i}^{(n)}(\bm{r}) u_i(\bm{r}) -\oint_{\partial \Omega}dA \sigma_{{\rm s} i} u_{i}.
\end{equation}
Here again the first term ${\cal F}_{\rm e}$ is the deformation energy to be discussed in Sec.~\ref{Sec:ActSolid-Elast}.  
In the second term, 
\begin{equation}\label{Eq:ActSolid-Wadef}
{\cal W}_{\rm a}\equiv \sum_{n=1}^{N} \int_{\Omega_n} d{\bm r}f_{{\rm a}i}^{(n)}(\bm{r}) u_i(\bm{r}) 
\end{equation} 
is the active work done by a \emph{discrete} distribution of active forces, originating mainly from ATP-consuming molecular motors, with volume density $\bm{f}_{\rm a}^{(n)}$ and $n=1,2,\ldots, N$. The third term is the work done by external forces with surface force density $\bm{\sigma}_{\rm s}$ acting as a Lagrange multiplier that enforces the boundary conditions at the system surfaces.


\begin{figure}[htbp]
  \centering
  \includegraphics[clip=true, viewport=1 1 400 300, keepaspectratio, width=0.35\textwidth]{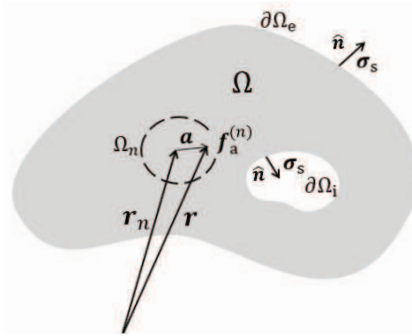}
  \caption {Schematic illustration of a continuum elastic solid with volume $\Omega$ bounded by the surface $\partial \Omega$ that includes the inner surface $\partial \Omega_{\rm i}$ and the outer surface $\partial \Omega_{\rm e}$. $\bm{\hat{n}}$ is the outward unit vector normal to the surface $\partial \Omega$. $\bm{\sigma}_{\rm s}$ is the stresses applied on the surface $\partial \Omega$. We also show typical active force density $\bm{f}_{\rm a}^{\rm (n)}$ that is distributed in a small volume $\Omega_{\rm n}$ around $\bm{r}_{\rm n}$.
  } \label{Fig:ActSolid-Volume}  
\end{figure}

As shown schematically in Fig.~\ref{Fig:ActSolid-Volume}, we consider the active force density $\bm{f}_{\rm a}^{(n)}$ distributed in a microscopic region $\omega_{\rm n}$ of dimension $a$ (a microscopic characteristic length such as the size of myosin motors) around $\bm{r}_{\rm n}$. The displacement vector $\bm{u}(\bm{r})$ around $\bm{r}_{\rm n}$ can be expanded in the microscopic region $\Omega_{\rm n}$ as $u_i(\bm{r}=\bm{r}_{\rm n}+\bm{a})=u_i(\bm{r}_n)+\sum_{m=1}^{\infty}\frac{1}{m!}a_{k_1}...a_{k_m}u_{i,k_1,...,k_m}(\bm{r}_n)$ with comma denoting the partial derivatives over spatial coordinates, and the active work ${\cal W}_{\rm a}$ in Eq.~(\ref{Eq:ActSolid-Wadef}) can then expanded to be 
\begin{equation}\label{Eq:ActSolid-Wa}
{\cal W}_{\rm a}=\int_{\Omega} d{\bm r} \sum_{m=1}^{\infty}\frac{1}{m!} 
\sum_{n=1}^{N} P_{k_1...k_mi}^{(n)} u_{i,k_1,...,k_m}(\bm{r})\delta(\bm{r}-\bm{r}_n). 
\end{equation} 
Here we have assumed the total (or net) active force is zero (since active forces are internal forces generated by molecular motors in biological systems): $\int_{\Omega_n} d{\bm r}f_{{\rm a}i}^{(n)}(\bm{r})=0$, and the multipole moment is defined as $P_{k_1...k_mi}^{(n)}=\int_{\Omega_n} d{\bm a}a_{k_1}...a_{k_m}f_{{\rm a}i}^{(n)}$ and particularly the active (contractile or extensile) force dipole moment tensor is
\begin{equation}\label{Eq:ActSolid-ForceDipole}
P_{ij}^{(n)}\equiv \int_{\Omega_n} d{\bm a}a_if_{{\rm a}j}^{(n)}.
\end{equation}
To the first (dipole) order, the total free energy is given by
\begin{equation}\label{Eq:ActSolid-Ft}
{\cal F}_{\rm t} = {\cal F}_{\rm e}+
\int_{\Omega} d{\bm r}{\sigma_{ij}^{\rm a}}\epsilon_{ij} -\oint_{\partial \Omega}dA \sigma_{{\rm s} i} u_{i},
\end{equation}
in which the active stress tensor
\begin{equation}\label{Eq:ActSolid-SigmaA0}
{\sigma_{ij}^{\rm a}}(\bm{r})=-\sum_{n=1}^{N} P_{ji}^{(n)} \delta(\bm{r}-\bm{r}_n),   
\end{equation} 
is introduced to physically represent the average density of active force dipoles applied on the elastic matrix. Minimization of ${\cal F}_{\rm t}$ with respect to $\bm{u}$ gives the equilibrium force balance equations and boundary conditions:
\begin{subequations}\label{Eq:ActSolid-EquilEqns12}
\begin{equation}\label{Eq:ActSolid-EquilEqns1}
\partial_j({\sigma_{ij}^{\rm e}}+{\sigma_{ij}^{\rm a}})=0, \quad \rm{in} \quad \Omega,
\end{equation}
\begin{equation}\label{Eq:ActSolid-EquilEqns2}
\hat{n}_j({\sigma_{ij}^{\rm e}}+{\sigma_{ij}^{\rm a}})=\sigma_{{\rm s} i},\quad {\rm or} \quad u_i=u_{{\rm s}i}, \quad \rm{at} \quad \partial \Omega.
\end{equation}
\end{subequations}
Note that the derivation is very similar to that in general elastic solids as shown in Eq.~(\ref{Eq:VarMeth-EquilEqn12}). 
The two conditions in Eq.~(\ref{Eq:ActSolid-EquilEqns2}) are denoted as natural and essential boundary conditions, respectively. Here $\sigma_{{\rm s}i}$ and $u_{{\rm s}i}$ are some constant stress and displacement at the boundary, respectively. The equation system in Eq.~(\ref{Eq:ActSolid-EquilEqns12}) should be supplemented by the constitutive relations for the stresses, $\bm{\sigma}^{\rm e}$ and $\bm{\sigma}^{\rm a}$, as functions of state variables of the active elastic solids. These constitutive relations will be discussed in the next two subsections.

\begin{figure}[htbp]
  \centering
  \includegraphics[clip=true, viewport=1 1 760 400, keepaspectratio, width=0.5\textwidth]{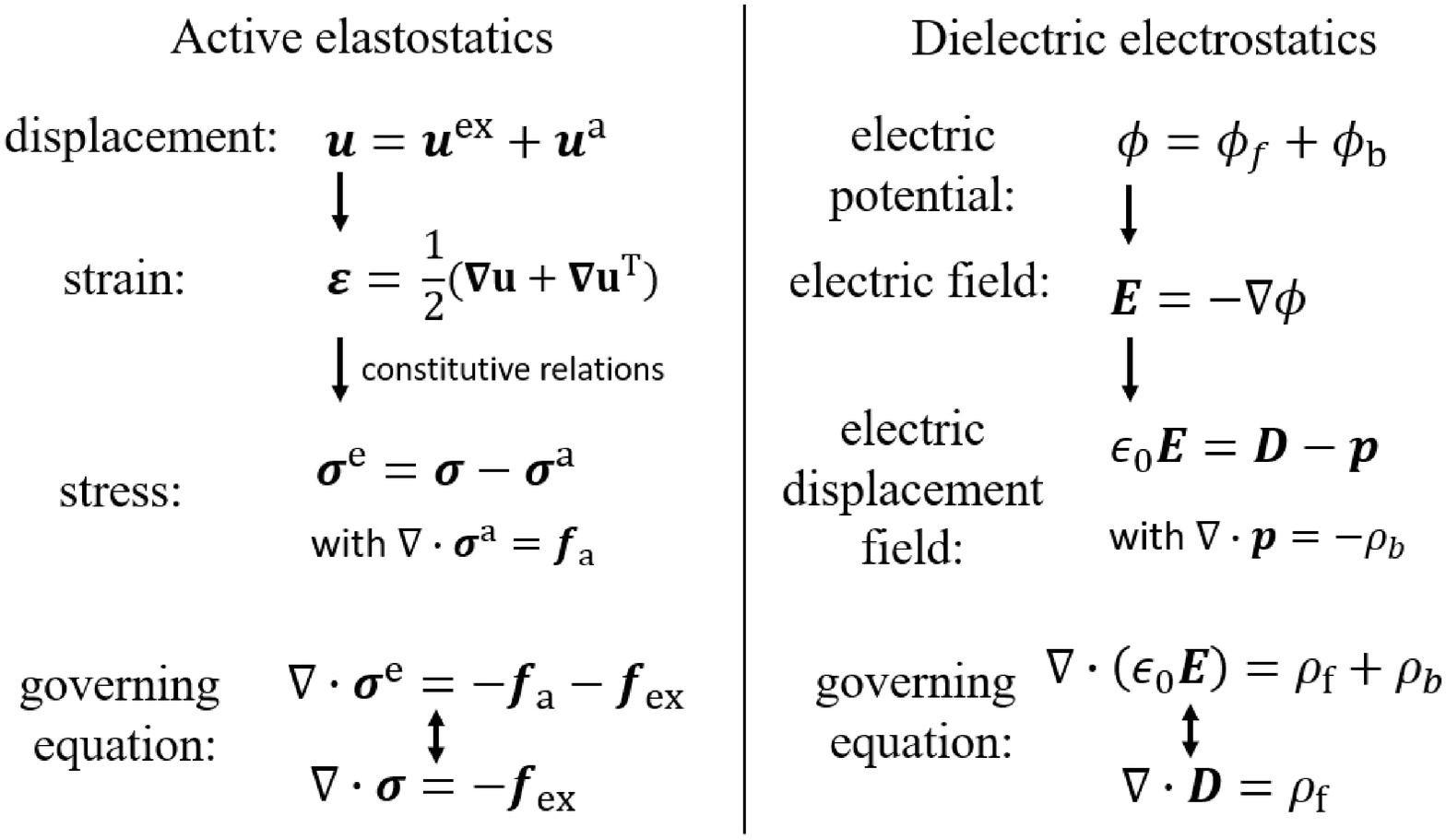}
  \caption {Formal analogy between elastostatics in elastic solids (in the presence of both active force dipoles and external forces with density $-\sigma^{\rm a}$ and $\bm{f}_{\rm ex}$, respectively) and electrostatics in dielectric media~\cite{Joanny2010} (in the presence of both permanent electric dipoles and free charges with density $\bm{p}$ and $\rho_{\rm f}$, respectively).
  }\label{Fig:ActSolid-Analogy}   
\end{figure}

To summarize the continuum theory of active elastic solids, it is interesting to note that there is a formal analogy between elastostatics in elastic solids in the presence of active force dipoles~\cite{Joanny2010,Sam2004,Sam2013a} and electrostatics in dielectric media with permanent electric dipoles (or ``frozen-in" polarization of polar molecules)~\cite{Purcell2013}. As shown in Fig.~\ref{Fig:ActSolid-Analogy}, the analogy is $u_{i} \leftrightarrow \phi$, $\epsilon_{ij} \leftrightarrow E_{i}$, $\sigma_{ij} \leftrightarrow D_{i}$, $\sigma_{ij}^{\rm a} \leftrightarrow p_{i}$, $f_{{\rm ex},i} \leftrightarrow-\rho_{\rm {f}}$, and $f_{{\rm a},i} \leftrightarrow-\rho_{\rm {b}}$.  
Here, $\phi$ is the total electric potential; $\phi_{\rm{f}}$ and $\phi_{\rm{b}}$ are the electric potentials induced by the free charges of density, $\rho_{\rm{f}}$, and the bound charges of density, $\rho_{\rm{b}}$, respectively. $E_i$ is the electric field and $D_{i}$ is the dielectric displacement with $\epsilon_{0}$ being the vacuum permittivity. The density of electric dipole, $p_{i}=\sum_{n} P_{i}^{(n)} \delta (\bm{r}-\bm{r}_{n})$ is the polarization. 


\subsection{Elastic stress: force dipole density of the matrix}\label{Sec:ActSolid-Elast}


We here summarize the constitutive relations of linear elastic materials that are represented by deformation energy~\cite{Landau1986,Lekhnitskii1981}.

Firstly, in linear isotropic elastic materials, the deformation energy density $F_{\rm e}$ is given by~\cite{Landau1986}
\begin{equation}\label{Eq:ActSolid-LinIsoFe}
F_{\rm e}=\frac{E_0}{2(1+\nu_0)} \left(\epsilon_{ij}^2 +\frac{\nu_0}{1-2 \nu_0} \epsilon_{kk}^2 \right)
\end{equation}
with $\epsilon_{ij}=\frac{1}{2}(\partial_i u_j+\partial_j u_i)$ being the strain components. Here, the elastic constants, $E_0$ and $\nu_0$, are Young's modulus and Poisson's ratio, respectively. From the deformation energy in Eq.~(\ref{Eq:ActSolid-LinIsoFe}) and using $\sigma_{ik}^{\rm e}=\partial F_{\rm e}/\partial \epsilon_{ik}$ (applicable also for nonlinear materials), we obtain the stress-strain relation or constitutive relation (Hooke's law)~\cite{Landau1986}:
\begin{equation}\label{Eq:ActSolid-LinIsoSigmaE}
{\sigma}_{ij}^{\rm e}= \frac{E_0}{1+\nu_0} \left(\epsilon_{ij} +\frac{\nu_0}{1-2\nu_0} \epsilon_{kk} \delta_{ij} \right).
\end{equation}

Next, we consider a typical linear anisotropic material -- transversely isotropic (or briefly \emph{transtropic}) material~\cite{Lekhnitskii1981}. A transtropic material is one with physical properties that are symmetric about an axis that is normal to a plane of isotropy, for example, hexagonal close-packed crystals~\cite{Landau1986} and nematic elastomers~\cite{Terentjev2007}. In this case, the deformation energy density is given, based on symmetry considerations, by~\cite{Landau1986,Lekhnitskii1981}
\begin{align}\label{Eq:ActSolid-LinAnisoFe}
F_{\rm e}= 
&\frac{1}{2}C_{1}\epsilon_{11}^2
+ 2C_{2}(\epsilon_{22}+\epsilon_{33})^2 +C_{3}[(\epsilon_{22}-\epsilon_{33})^2+4\epsilon_{23}^2]\\ \nonumber
&+2C_{4}\epsilon_{11}(\epsilon_{22}+\epsilon_{33})
+2C_{5}(\epsilon_{12}^2+\epsilon_{13}^2),
\end{align} 
where we have taken $\hat{\bm{r}}_1$ as the axis of symmetry, $\hat{\bm{r}}_2$ and $\hat{\bm{r}}_3$ span the plane of isotropy. Note that for transtropic materials, there are \emph{five} independent elastic constants, $C_i$ ($i=1,...,5$), which relate to Young's moduli and Poisson's ratios in the form 
\begin{equation}\label{Eq:ActSolid-LinAnisoC}
\begin{aligned}
C_{1}&=\frac{E_{1}}{m}(1-\nu_{23}),  \quad
C_{2}=\frac{E_{2}}{8m}, \quad
C_{3}=\frac{1}{2}\mu_{23}, \\
C_{4}&=\frac{E_{2}\nu_{12}}{2m}, \quad\quad
C_{5}=\mu_{12},  
\end{aligned}
\end{equation}
with $m\equiv 1-\nu_{23}-2\nu_{12}^2E_2/E_1$. Here $E_1$ and $E_2$ are the Young's moduli along the $\hat{\mathbf{x}}_1$-axis of symmetry and in the isotropic ($\hat{\mathbf{x}}_2$--$\hat{\mathbf{x}}_3$) plane, respectively.
$\nu_{ij} \equiv - {\partial \epsilon_j}/{\partial \epsilon_i}$  (with $i\neq j$ and $i,j=1,2,3$) are the differential Poisson's ratios for tensile stress applied along $i$-direction and contraction in $j$-direction, following the general notation in anisotropic materials\cite{Lekhnitskii1981}. $\mu_{12}$, $\mu_{23}$ are the shear moduli in the $\hat{\mathbf{x}}_1$--$\hat{\mathbf{x}}_2$ plane and the isotropic $\hat{\mathbf{x}}_2$--$\hat{\mathbf{x}}_3$ plane, respectively. From the free energy in Eq.~(\ref{Eq:ActSolid-LinAnisoFe}), we obtain the stress-strain relations 
\begin{subequations}\label{Eq:ActSolid-LinAnisoSigmaE}
\begin{align}
   \sigma_{11}^{\rm e}&= C_{1}\epsilon_{11} +2C_{4}\epsilon_{22} +2C_{4}\epsilon_{33}\\
   \sigma_{22}^{\rm e}&= 2C_{4}\epsilon_{11} +(4C_{2}+2C_{3})\epsilon_{22} +(4C_{2}-2C_{3})\epsilon_{33}\\
   \sigma_{33}^{\rm e}&= 2C_{4}\epsilon_{11} +(4C_{2}-2C_{3})\epsilon_{22} +(4C_{2}+2C_{3})\epsilon_{33}\\
   \sigma_{12}^{\rm e}&= 2C_{5}\epsilon_{12},\quad \sigma_{13}^{\rm e}=2C_{5}\epsilon_{13},\quad \sigma_{23}^{\rm e}=2C_{5}\epsilon_{23},
\end{align}
\end{subequations}

\subsection{Active stress: force dipole density applied by active units}\label{Sec:ActSolid-Act}

\begin{figure}[htbp]
  \centering
  \includegraphics[width=0.47\textwidth]{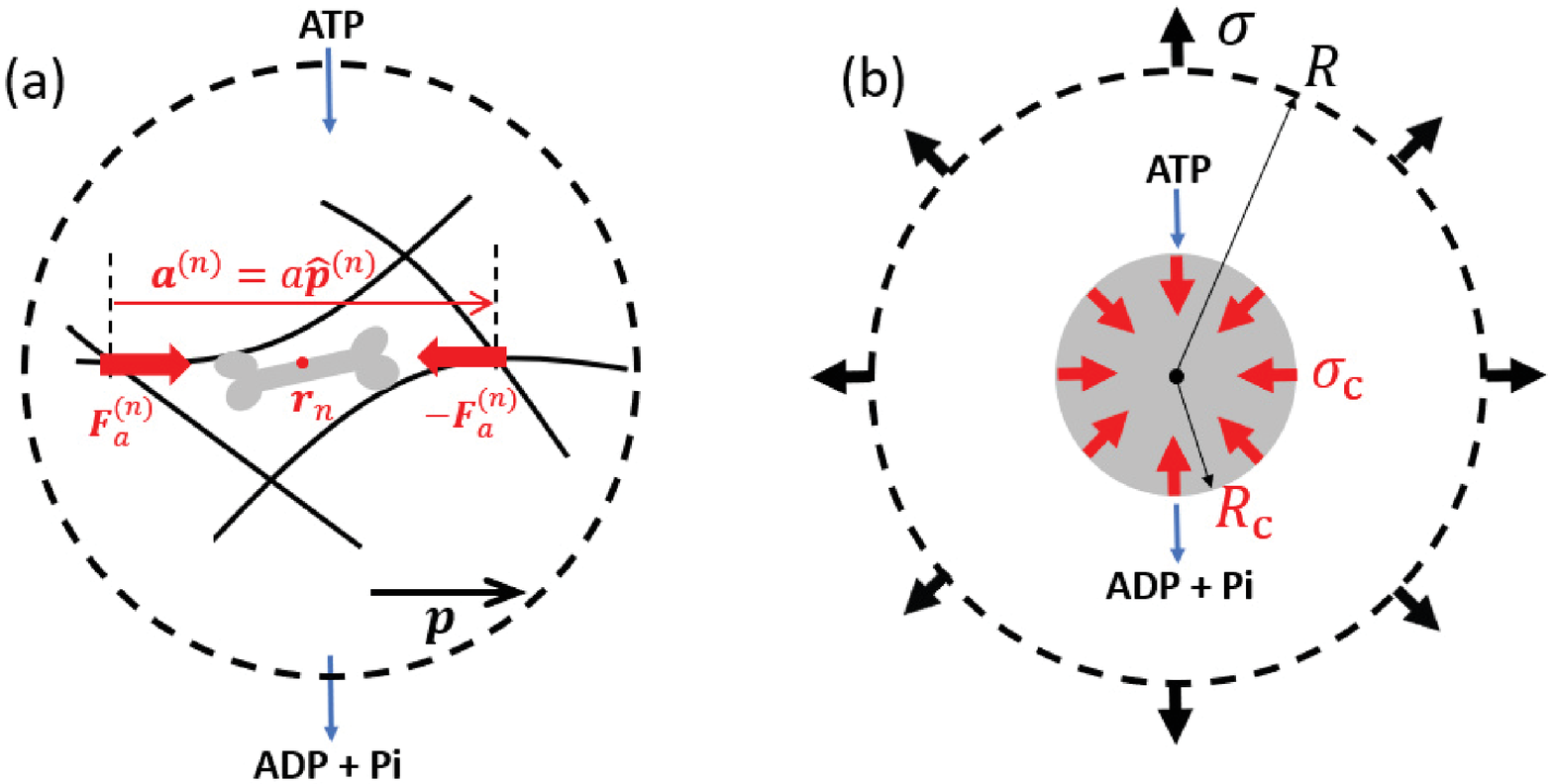}
  \caption {Schematic illustration of a representative volume element (RVE) in active elastic solids related to cell mechanics where the concept of force dipoles is found to be useful at multiple scales. (a) At subcellular scales, where actomyosin complexes induce active contractile force dipoles in the cytoskeleton. (b) At cellular scales, adherent cells generate contractile active force dipoles in its surrounding extracellular matrix. In both cases, active stresses are calculated and both found to be proportional to the density of active force dipoles.
  }\label{Fig:ActSolid-Act}  
\end{figure}

Active stress is a key feature distinguishing active matter from inert or passive matter. In the biological context, active stress originate from molecular processes that are often mediated by molecular motors. These motors are fueled by the hydrolysis reaction of adenosine triphosphate (ATP) to adenosine diphosphate (ADP) and inorganic phosphate ($\mathrm{P_i}$)~\cite{Alberts2007}, as shown in Fig.~\ref{Fig:ActSolid-Act}. This hydrolysis reaction is therefore the dominant chemical reaction that couples via the coefficient $\zeta$ to active stress. Because of Onsager symmetries of the coupling coefficients, the coefficients $\zeta$ also enter the chemical reaction rates, which effectively become mechanosensitive. Here we calculate the active stress in two specific biological scenarios at different length scales: (i) active point force dipoles, related to actomyosin cytoskeleton at sub-cellular scales; (ii) active spherical force dipoles, related to cell-matrix composites at cellular scales.



\emph{Active point force dipoles.} Consider the cytoskeleton of adherent animal cells, in which a small assembly of myosin motors bind transiently to the cross-linked actin network, consume ATP and locally contract the network, generating pairs of equal and opposite forces $\pm \bm{F}^{(n)}=\pm F^{(n)} \hat{\bm{p}}^{(n)}$ (see Fig.~\ref{Fig:ActSolid-Act}(a)). Here $\hat{\bm{p}}^{(n)}$ are unit vectors, characterizing the opposite orientations of the two aligned filaments and $n$ is an index over such pairs of aligned filaments that are cross-linked by active motors. These force pairs correspond to active force dipoles with force density per volume
\begin{equation}\label{Eq:ActSolid-SigmaAfa1}
f_{\rm{a}i}^{(n)}=F^{(n)} \hat{p}_{i}^{(n)}\left[\delta\left(\bm{r}-\bm{r}_{n}+\frac{1}{2} \bm{a}^{(n)}\right)-\delta\left(\bm{r}-\bm{r}_{n}-\frac{1}{2} \bm{a}^{(n)}\right)\right]
\end{equation}
at position $\bm{r}_{n}$ in the network. Here $a$ is a microscopic length (as mentioned previously), here representing the size of myosin motors (or myosin filaments), or specifically the distance at which the pair forces act. On large length scales or in a continuum limit, this can be expanded in the microscopic length $a$. To first order it corresponds to the point force dipole
\begin{equation}\label{Eq:ActSolid-SigmaAfa2}
f_{\rm{a}i}^{(n)} \simeq - \partial_{j} P_{ji}^{(n)} \delta\left(\bm{r}-\bm{r}_{n}\right)
\end{equation}
with the dipole moment 
$P_{ij}^{(n)}\equiv -aF^{(n)} \hat{p}_{i}^{(n)} \hat{p}_{j}^{(n)}=P_{ji}^{(n)}$. 
In a continuum description, the density of these active point force dipoles in a local volume elements $V$ is the active stress
\begin{equation}\label{Eq:ActSolid-SigmaA1}
\sigma_{ij}^{\rm{a}}=-\frac{1}{V} \sum_{n} P_{ij}^{(n)}=\frac{1}{V} \sum_{n} aF \hat{p}_{i}^{(n)} \hat{p}_{j}^{(n)}=-\rho P \hat{p}_{i}\hat{p}_{j}
\end{equation}
where the sum is over motor-induced force dipoles in the volume $V$ and the local averaged polarization vector is given by $\hat{p}_{i}=\frac{1}{N_{\rm P}}\sum_{n=1}^{N_{\rm P}} \hat{p}_{i}^{(n)}$ with $N_{\rm P}$ being the total number of force dipoles in the volume $V$. It is important to note that for contractile force dipoles, $F$ is positive and hence the principal $\hat{\bm{p}}^{(n)}\hat{\bm{p}}^{(n)}$ component of the active force dipole moment is negative.

\emph{Active spherical force dipoles.} Now let's consider adherent cells that are embedded and contracting in elastic extracellular matrix. If the cells are contracting in a needle-like manner, the above discussions using point force dipoles still apply. However, here we examine a special cell-matrix composite with spherically contracting cells as shown in Fig.~\ref{Fig:ActSolid-Act}(b). In a spherical unit cell or bounding volume of radius $R$, a single spherical cell of radius $R_{\rm c}$ is contracting the elastic matrix by applying a stress $\sigma_{\rm c}$ on its boundary, $r=R_{\rm{c}}$. In this specific simple case with spherical symmetry, the active stress can be calculated explicitly as follows~\cite{Martin2019}. 

The dipole moment tensor (defined in Eq.~(\ref{Eq:ActSolid-ForceDipole})) of the spherically symmetric force dipole is isotropic and reads $P_{i j}=P\delta_{i j}$ with $P=-\frac{4}{3}\pi \sigma_{\rm c} R_{\rm c}^3$. To calculate the active stress, we further assume that a constant external stress $\sigma_{r r}(R)=\sigma$ is applied at the outer boundary ($r=R$) of the bounding volume where $\sigma$ may be positive or negative. If the matrix is linear and generally we assume it to be anisotropic, the total free energy is given by\cite{Landau1986}
\begin{equation}\label{Eq:ActSolid-SigmaAFt}
{\cal F}_{\rm t}= \int_0^R [F_{\rm e}+ \sigma_{\rm c}\delta(r-R_{\rm c})]4\pi r^2dr -\sigma4\pi R^2 u(R),
\end{equation} 
in which the elastic energy density is given by $F_{\rm e}= 
\frac{1}{2}C_{1}u'^2
+ 8C_{2}\frac{u^2}{r^2}+4C_{4}\frac{uu'}{r}$.
Minimizing ${\cal F}_{\rm t}$ in Eq.~(\ref{Eq:ActSolid-SigmaAFt}) with respect to $u(r)$ gives the equilibrium equations and boundary conditions:
\begin{subequations}\label{Eq:ActSolid-SigmaEqns12}
\begin{equation}\label{Eq:ActSolid-SigmaEqns1}
    \frac{d^2u}{dr^2} +\frac{2}{r}\frac{du}{dr} -{\cal B}\frac{2u}{r^2} =0,
\end{equation}
\begin{equation}\label{Eq:ActSolid-SigmaEqns2}
\lim_{\varepsilon\to 0}\left[\sigma_{r r}(r=R_{\rm c}+\varepsilon)-\sigma_{r r}(r=R_{\rm c}-\varepsilon)\right]=\sigma_{\rm c}, \quad \sigma_{r r}(r=R)=\sigma.
\end{equation}
\end{subequations}
Here the dimensionless parameter is 
\begin{equation}\label{Eq:ActSolid-B}
{\cal B}\equiv (8C_2-2C_4)/C_1,  
\end{equation}
and using Eq.~(\ref{Eq:ActSolid-LinAnisoC}) we get ${\cal B}= {E_{2}(1-\nu_{12})}/{E_{1}(1-\nu_{23})}$ with $\nu_{12}$ and $\nu_{23}$ being the Poisson ratios in the radial and transverse planes, respectively, The stress component $\sigma_{\rm rr}$ is
$\sigma_{\rm rr}=C_1 u' + 4C_4 {u}/{r}$. 
Solving this 1D boundary value problem, we obtain the displacement field:
\begin{subequations}\label{Eq:ActSolid-SigmaASol12}
\begin{equation}\label{Eq:ActSolid-SigmaASol1}
\begin{aligned}
&u(r\leq R_{\rm c})= \left(\frac{r}{R}\right)^{n-1} \left\{\frac{\sigma R}{(n-1)C_1+4 C_4} -\right.\\
&\left. \frac{\sigma_{\rm c} R}{(2n-1)C_1}\left(\frac{R}{R_{\rm c}}\right)^{n-2}\left[1+\left(\frac{R_{\rm c}}{R}\right)^{3}\frac{nC_1-4 C_4}{(n-1)C_1+4 C_4}\right]\right\} .
\end{aligned}
\end{equation}
\begin{equation}\label{Eq:ActSolid-SigmaASol2}
\begin{aligned}
&u(r\geq R_{\rm c})= \frac{\sigma R}{(n-1)C_1+4 C_4} \left(\frac{r}{R}\right)^{n-1}-\frac{\sigma_{\rm c} R}{(n-1)C_1+4 C_4}  \\
& \left(\frac{R_{\rm c}}{R}\right)^{n+1}\left[\frac{nC_1-4 C_4}{ (2n-1)C_1}  \left(\frac{r}{R}\right)^{n-1} +\frac{(n-1)C_1+4 C_4}{(2n-1)C_1}\left(\frac{R}{r}\right)^{n}\right].
\end{aligned}
\end{equation}
\end{subequations} 
with $n$ given by
\begin{equation}\label{Eq:ActSolid-SigmaAn}
n=\frac{1}{2}\left(1+\sqrt{1+8{\cal B}}\right).
\end{equation}
Particularly at $r=R$, we have
\begin{equation}\label{Eq:ActSolid-SigmaASoluR}
u(R)= \frac{R}{(n-1)C_1+4 C_4}\left[\sigma-\sigma_{\rm c} \left(\frac{R_{\rm c}}{R}\right)^{n+1}\right],
\end{equation}
which describes the deformation (or the total displacement) of the bounding volume that is induced by both external stress and internal active stress.

The external stress $\sigma$ applied at the boundary of the elastic matrix is balanced by two contributions: a passive elastic response of the network, and an active stress specifically due to the presence of these active units
\begin{equation}\label{Eq:ActSolid-SigmaEA}
\sigma=\sigma^{\rm{e}}+\sigma^{\rm{a}}.
\end{equation}
The elastic stress $\sigma^{\rm{e}}$ can be determined as the stress required at the outer boundary $r=R$ to produce the same deformation (as given in Eq.~(\ref{Eq:ActSolid-SigmaASoluR})) of a purely passive bounding volume (with $\sigma_{\rm{c}}=0$):
\begin{equation}\label{Eq:ActSolid-SigmaAE}
\sigma^{\rm{e}}=\left[(n-1)C_1+4 C_4)\right] \frac{u(R)}{R}.
\end{equation} 
Thus, the active stress is proportional to the force dipole and density of active units through  
\begin{equation}\label{Eq:ActSolid-SigmaA2}
\sigma_{ij}^{\rm{a}}=\sigma_{\rm{c}}\left(\frac{R_{\rm{c}}}{R}\right)^{n+1}\delta_{ij}=-\rho P\phi^{(n-2)/3}\delta_{ij}.
\end{equation}
in which $\rho=({4\pi R^3}/{3})^{-1}$ and $\phi=R_{\rm{c}}^3/{R}^3$ are the number density and the volume fraction of the contractile cells or force dipoles, respectively. Particularly, if the matrix is linear isotropic, then ${\cal B}=1$ and hence $n=2$, the active stress reads~\cite{Martin2019} $\sigma_{ij}^{\rm{a}}=-\rho P\delta_{ij}$. If $E_2 \ll E_1$ and hence $n=1$ from Eq.~(\ref{Eq:ActSolid-SigmaAn}), the active stress is given by $\sigma_{ij}^{\rm{a}}=-\rho P \phi^{-1/3}\delta_{ij}$, which is denoted as the ``density-controlled" regime corresponding to the fully buckled networks~\cite{Martin2019}. Moreover, it has also been shown~\cite{Martin2019} that the active stresses can be amplified significantly by the nonlinear elasticity of the biopolymer matrix. 

In summary, the active stress $\sigma_{\alpha \beta}^{\text {a}}$ emerges from a large number of force generating events that occur generally in the anisotropic material. In a coarse-grained continuum description, the active stress in linear elastic solids can be generally decomposed into the following form~\cite{Prost2015}
\begin{equation}\label{Eq:ActSolid-SigmaA}
\sigma_{i j}^{\rm a}=-\tilde{\zeta} q_{i j}-\bar{\zeta} \delta_{i j},
\end{equation}
in which $q_{i j}=(\hat{p}_{i} \hat{p}_{j}-\frac{1}{3} \delta_{i j})$ is the nematic order parameter with $\hat{p}_{i}$ being a unit vector in the direction of the filaments. 
The first term is traceless and anisotropic, while the second term is isotropic. 
The coefficients $\tilde{\zeta}$ and $\bar{\zeta}$ depend on motor and filament densities, and vanish when the difference $\Delta \mu$ between the chemical potential of the fuel (ATP) and that of the reaction products vanishes. The active stress is contractile (extensile) along the nematic or polar axis if $\tilde{\zeta}$ and $\bar{\zeta}$ are negative (positive). In experiments on the cell cortex and lamellipodia~\cite{Eva2009,Prost2015}, the active stress is found to be the order of $0.1$ to a few $\rm kPa$, depending on biochemical regulation.  
In addition, we would like to point out that although the above discussions on active stresses are made in the context of active elastic solids, most of them also apply to active fluids, for example, active stress as density of force dipoles and equilibrium or steady states determined by the balance between active stresses and passive stresses, \emph{etc}.

\section{Spontaneous bending and contraction of active circular plates}\label{Sec:App1}

Now we use the above variational methods and the deep Ritz method to study the bending and contraction of a circular plate that is induced by asymmetric active contractile stresses with and without gravitational effects. This simple model problem is related to the morphology of several biology systems at multiple length scales such as \emph{in vitro} reconstituted actomyosin gels, individual contractile adherent cells, and a solid-like monolayer or bilayer of confluent adherent cells. 


\subsection{General equilibrium equations in cylindrical coordinates}\label{Sec:App1-Eqns}

Consider a circular elastic plate of thickness $h$ and lateral radius $R$ in the presence of active stresses and suspended by its center (see Fig.~\ref{Fig:App1-Bending}(a)). We assume that the distribution of active stresses has axial rotational symmetry, hence the resulted deformation has cylindrical symmetry. In this case, we therefore take cylindrical polar coordinates ($r,\theta,z$) as schematically shown in Fig.~\ref{Fig:App1-Bending}(a), in which the displacement vector is given by $\bm{u}=\left(u_r(r,z), u_\theta =0, u_z(r,z) \right)$ and the non-zero components of the strain tensor $\epsilon_{ij}$ at small deformations are
\begin{equation}\label{Eq:App1-Strain}
\begin{aligned} 
&\epsilon_{\rm rr} = \partial_{\rm r} u_{\rm r}, \quad 
\epsilon_{\rm \theta \theta} = u_{\rm r}/r, \quad 
\epsilon_{\rm zz} = \partial_{\rm z} u_{\rm z}\\
&\epsilon_{\rm rz} = \epsilon_{\rm zr}=\frac{1}{2}\left(\partial_{\rm z} u_{\rm r}+\partial_{\rm r} u_{\rm z}\right).     
\end{aligned}    
\end{equation}
Furthermore, we assume the active stresses apply only on the $r-\theta$ plane and take the following in-plane isotropic form~\cite{Joanny2009}: 
\begin{equation}\label{Eq:App1-SigmaA}
\bm{\sigma}^{\rm a}=-\zeta(\hat{\bm{r}}\hat{\bm{r}}+\hat{\bm{\theta}}\hat{\bm{\theta}}).
\end{equation}
In general, $\bm{\sigma}^{\rm a}$ is non-uniform along the radial direction and asymmetric through the plate thickness, that is, $\zeta=\zeta(r, z)$ ($<0$ for contraction), and here we consider the following simple step-wise form (as schematically shown in Fig.~\ref{Fig:App1-Bending}(a,b))
\begin{align}\label{Eq:App1-Zeta}
\zeta = \zeta_{\rm a} \Theta(z-(1-2 \alpha)\frac{h}{2}),  
\end{align}
with $\zeta_{\rm a}=\zeta_0<0$ and $\zeta_{\rm a}=\zeta_0 r/R<0$ for uniform and non-uniform contraction, respectively, and $\Theta$ being Heaviside step function. Here $\alpha=h_{\rm a}/h \in [0,1]$ measures the fraction of active layer (with thickness $h_{\rm a}$): $\alpha=0$ representing a purely passive plate, and $\alpha=1$ representing a symmetric active plate with uniform active stresses along the thickness direction.

For linear isotropic materials with non-zero strain components given in Eq.~(\ref{Eq:App1-Strain}), the total energy is obtained from Eq.~(\ref{Eq:ActSolid-Ft}) and Eq.~(\ref{Eq:ActSolid-LinIsoFe}) as
\begin{equation}\label{Eq:App1-Ft}
\begin{aligned}
{\cal F}_{\rm t} = 
& \int_{-h/2}^{h/2} dz \int_0^R dr 2 \pi r \left\{\frac{E_0}{2(1+\nu_0)}\left[ \left(\epsilon_{\rm rr}^2+\epsilon_{\rm \theta\theta}^2 +\epsilon_{\rm zz}^2+2 \epsilon_{\rm rz}^2\right) \right.\right.\\
&\left.\left. +\frac{\nu_0}{1-2\nu_0}\left(\epsilon_{\rm rr} +\epsilon_{\rm \theta\theta} +\epsilon_{\rm zz}\right)^2\right]+\rho g u_{\rm z}-\zeta (\epsilon_{\rm rr}+\epsilon_{\rm \theta\theta})\right\}. 
\end{aligned}
\end{equation} 
Here the body force $\rho g$ corresponds to the direct effect of gravity plus Archimedes buoyancy force~\cite{Joanny2021}.
Substituting the strain components in Eq.~(\ref{Eq:App1-Strain}) and minimizing ${\cal F}_{\rm t}$ with respect to $u_{\rm r}$ and $u_{\rm z}$  (the same as how we have obtained Eq.~(\ref{Eq:VarMeth-EquilEqn12})) give the bulk equilibrium conditions
\begin{subequations}\label{Eq:App1-EquilEqns12}
\begin{equation}\label{Eq:App1-EquilEqns1}
\begin{aligned}
&(\nabla \cdot \bm{\sigma})_{\rm r}=\frac{1}{r}\frac{\partial (r \sigma_{\rm rr})}{\partial r}+\frac{\partial \sigma_{\rm rz}^{\rm e}}{\partial z}-\frac{1}{r}\sigma_{\rm \theta\theta}=0, \\
&(\nabla \cdot \bm{\sigma})_{\rm z}=\frac{1}{r}\frac{\partial (r \sigma_{\rm zr}^{\rm e})}{\partial r}+\frac{\partial \sigma_{\rm zz}^{\rm e}}{\partial z}=\rho g,
\end{aligned}
\end{equation}
and natural boundary conditions at $rR$ and $z=\pm h/2$, respectively, as 
\begin{equation}\label{Eq:App1-EquilEqns2}
\begin{aligned}
&\sigma_{\rm rr}|_{r=R}=\sigma_{\rm r\theta}|_{r=R}=\sigma_{\rm rz}^{\rm e}|_{r=R}=0, \\
&\sigma_{\rm zz}^{\rm e}|_{z=\pm h/2}=\sigma_{\rm zr}^{\rm e}|_{z=\pm h/2}=\sigma_{\rm z\theta}^{\rm e}|_{z=\pm h/2}=0,
\end{aligned}
\end{equation} 
\end{subequations} 
which are supplemented with conditions at $r=0$: $u_{\rm r}(r=0,z)=0$ and $u_{\rm z}(r=0,z=0)=0$. Here the total stress tensor is $\bm{\sigma}=\bm{\sigma}^{\rm e} + \bm{\sigma}^{\rm a}$ with the active stress $\bm{\sigma}^{\rm a}$ given in Eq.~(\ref{Eq:App1-SigmaA}) and the elastic stress $\bm{\sigma}^{\rm e}$ given in Eq.~(\ref{Eq:ActSolid-LinIsoSigmaE}) and $\sigma_{\rm r \theta}^{\rm e}=\sigma_{\rm \theta r}^{\rm e}=\sigma_{\rm z \theta}^{\rm e}=\sigma_{\rm \theta z}^{\rm e}=0$. 

Note that the complete equation system in Eq.~(\ref{Eq:App1-EquilEqns12}) applies generally for the small deformations of linear elastic materials with isotropic in-plane active stresses in any cylindrical geometry with rotational axial symmetry. Their analytical solutions are available only in some simplest cases~\cite{Timoshenko1959,Bower2009}, but they can be easily solved numerically using some commercial finite element software. Here we will not pursue these solutions in general complex situations, but we instead focus on their solutions in thin-plate limits. 


For a thin circular plate, its thickness $h$ is much smaller than their lateral radius $R$. The normal stresses at the surfaces $z=\pm h/2$ are zero~\cite{Landau1986,Bower2009}: $\sigma_{i{\rm z}}=\sigma_{i{\rm z}}^{\rm e}=0$, where $\sigma_{i{\rm z}}^{\rm a}=0$ because the active stress has been assumed to be present only in the $r-\theta$ plane as shown in Eq.~(\ref{Eq:App1-SigmaA}). Then from the stress-strain relation in Eq.~(\ref{Eq:ActSolid-LinIsoSigmaE}), we have $\epsilon_{\rm rz}=0$ and $\epsilon_{\rm zz} = -\frac{\nu_0}{1-\nu_0} (\epsilon_{\rm rr}+\epsilon_{\rm \theta\theta})$. Therefore, the deformation energy in Eq.~(\ref{Eq:App1-Ft}) reduces to 
\begin{equation}\label{Eq:App1-ThinFt}
\begin{aligned}
{\cal F}_{\rm t} = &\int_{-h/2}^{h/2} dz \int_0^R dr 2 \pi r \left[\frac{E_0}{2(1-\nu_0^2)} \left(\epsilon_{\rm rr}^2+\epsilon_{\rm \theta\theta}^2 +2\nu_0 \epsilon_{\rm rr}\epsilon_{\rm \theta\theta}\right)\right.\\
&\left.+\rho g u_{\rm z}-\zeta (\epsilon_{\rm rr}+\epsilon_{\rm \theta\theta})\right]. 
\end{aligned}
\end{equation} 
In this case, the in-plane and out-plane deformation can be considered by using the classical Kirchhoff's plate theory~\cite{Timoshenko1959,Bower2009,Masters1994,Freund2000} and assuming the displacement components to be the simple form of 
\begin{equation}\label{Eq:App1-Displacements}
u_{\rm r} = u(r)-zw', \quad
u_{\rm z} \approx w(r),
\end{equation}
with the prime hereafter denoting the ordinary derivatives, in which $u(r)$ and $w(r)$ represent the in-plane displacement and the out-plane deflection of the neutral surface, respectively. 

\subsection{Variational formulations and Ritz method of approximation}\label{Sec:App1-Thin}

\subsubsection{Spontaneous bending at small deflections}\label{Sec:App1-Thin-Eqns}

We first consider the case of small deflections with $w\ll h$ and no gravity $\rho g=0$. In this case, we substitute the displacement components in Eq.~(\ref{Eq:App1-Displacements}) into Eq.~(\ref{Eq:App1-Strain}) and obtain the strain components: 
\begin{equation}\label{Eq:App1-ThinStrains}
\epsilon_{\rm rr}=u'-zw'', \quad
\epsilon_{\rm \theta\theta}=u/r-zw'/r.
\end{equation}
Substituting them into Eq.~(\ref{Eq:App1-ThinFt}) and integrating over thickness $z$-direction, we obtain the total energy as a functional of $u(r)$ and $w(r)$:
\begin{align}\label{Eq:App1-ThinFtuw}
{\cal F}_{\rm t}[u(r),w(r)] &= \int_0^R dr 2\pi r \left[\frac{Y}{2} \left(u'^2 +u^2/r^2+2\nu_0 uu'/r\right)\right. \nonumber \\
&\left.- \tau_{\rm a}(u'+u/r)+ \frac{D}{2}\left(w''^2 +w'^2/r^2+2\nu_0 w'w''/r\right)\right. \nonumber \\
&\left.-M_{\rm a}(w''+w'/r)\right].
\end{align} 
in which $Y=E_{0}h/(1-\nu_{0}^2)$ is the compression modulus, $\tau_{\rm a}=\alpha \zeta_{\rm a} h$, $D=E_0h^3/12(1-\nu_0^2)$ is the flexural stiffness or rigidity of the thin plate~\cite{Timoshenko1959,Bower2009,Masters1994,Freund2000}, and $M_{\rm a}=-\alpha(1-\alpha) \zeta_{\rm a}h^2/2$ is the bending moment applied by isotropic active stress $\bm{\sigma}^{\rm a}$ in Eq.~(\ref{Eq:App1-SigmaA}).
The first variation of ${\cal F}_{\rm t}$ is then given by
\begin{equation}\label{Eq:App1-DeltaFt}
\begin{aligned}
\delta{\cal F}_{\rm t}=& \int_0^R dr 2\pi r  \left[-h\overline{(\nabla \cdot \bm{\sigma})_{\rm r}} \delta u+(\rho gh-\nabla \cdot (\nabla \cdot \bm{M}))\delta w \right] \\
&+ \left[2\pi r  h\overline{\sigma}_{\rm rr} \delta u +2\pi r \hat{\bm r}\cdot (\nabla \cdot \bm{M})\delta w -2 \pi r M_{\rm rr} \delta w' \right]_0^R.
\end{aligned}
\end{equation}
Here the upper bar $\overline{(\ldots)}$ denotes the average over the plate thickness ($z$-direction).
The internal bending moment tensor $\bm{M}$ is given by $\bm{M}=[-D(w''+\nu_0 w'/r)+M_{\rm a}]\hat{\bm{r}}\hat{\bm{r}}+[-D(\nu_0 w'' +w'/r)+M_{\rm a}]\hat{\bm{\theta}}\hat{\bm{\theta}}$ in which $M_{\alpha\beta}$ denotes the bending moment generated by stress $\sigma_{\alpha\beta}$.
The term $\hat{\bm r}\cdot (\nabla \cdot \bm{M})=-\left(D\Delta w - M_{\rm a}\right)'$ is the Kirchhoff's effective shear force per unit length of the plate applied at the plate edge. $\Delta$ is the Laplace operator given in cylindrical coordinates by $\Delta=\frac{1}{r}\frac{d}{dr}\left(r\frac{d}{dr}\right)$. For free plate edges, Kirchhoff's edge conditions apply, consisting of a vanishing normal moment and a vanishing Kirchhoff's effective shear force. 

Then from the minimization of ${\cal F}_{\rm t}$ with respect to $u(r)$ and $w(r)$ gives the equilibrium equations (\emph{i.e.}, the Euler-Lagrange equations) in the bulk (for $0\leq r\leq R$) as
\begin{subequations}\label{Eq:App1-ThinEqns12345}
\begin{equation}\label{Eq:App1-ThinEqns1}
Y\left(u''+\frac{u'}{r}-\frac{u}{r^2}\right) -\tau_{\rm a}=0,
\end{equation}
\begin{equation}\label{Eq:App1-ThinEqns2}
D\left( w^{(4)}+\frac{2w'''}{r}-\frac{w''}{r^2}+\frac{w'}{r^3}\right)
-\left( M''_{\rm a}+\frac{M'_{\rm a}}{r} \right)=-\rho gh,
\end{equation} 
and boundary conditions at $r=R$, respectively, as
\begin{equation}\label{Eq:App1-ThinEqns3}
\left[Y\left(u'+\nu_0 \frac{u}{r}\right)-\tau_{\rm a} \right]_{r=R}=0, 
\end{equation}
\begin{equation}\label{Eq:App1-ThinEqns4}
\left[D\left(w'''+\frac{w''}{r}-\frac{w'}{r^2}\right)-M'_{\rm a}\right]_{r=R}=0,
\end{equation}
\begin{equation}\label{Eq:App1-ThinEqns5}
\left[D(w''+\nu_0\frac{w'}{r})-M_{\rm a}\right]_{r=R}=0,
\end{equation}
\end{subequations}
which are supplemented with conditions at $r=0$: $u=w=0$ and $w'=0$.
Here $M_{\rm a}=-\alpha(1-\alpha) \zeta_{\rm a}h^2/2$ that $\zeta_{\rm a}$ defined in Eq.~(\ref{Eq:App1-Zeta}) is the component of the active stress.
Note that the equation system in Eq.~(\ref{Eq:App1-ThinEqns12345}) can also be derived directly from the general equation system in Eq.~(\ref{Eq:App1-EquilEqns12}) obtained in the previous subsection.


For uniform and the simple non-uniform active stresses discussed in Eq.~(\ref{Eq:App1-Zeta}), the above equation system can be solved analytically as follows.
\begin{itemize}
    \item \emph{Uniform active contraction with $\zeta_{\rm a}=\zeta_0$.} The exact analytical solution is
    \begin{equation}\label{Eq:App1-ThinSol1}
    u(r)=-\frac{1}{6}\alpha{\cal Z}_{\rm b}\frac{h^2}{R^2}r, \quad
    w(r)=\frac{1}{2} \alpha(1-\alpha) {\cal Z}_{\rm b}\frac{h}{R^2} r^2,
    \end{equation} 
    in which the normalized parameter
    \begin{equation}\label{Eq:App1-Zcb}
        {\cal Z}_{\rm b}\equiv -{\zeta_0 h R^2}/{2D(1+\nu_0)}
    \end{equation}    
    measure the strength of active contractility in contracting (in-plane) and bending (out-of-plane) the plate, respectively. For active contraction with $\zeta_0<0$ and hence ${\cal Z}_{\rm b}>0$. From Eq.~(\ref{Eq:App1-ThinSol1}), we obtain that the two principal bending curvatures are equal and take the dimensionless form of  $w''h=w'h/r=\alpha(1-\alpha) {\cal Z}_{\rm b}h^2/R^2$, which is a uniform constant through the whole circular plate. Therefore, the bending curvatures are positive constants, which means the plate is bent up (toward the direction of active contracting layer) to be a spherical cap as schematic in Fig.~\ref{Fig:App1-Bending}(a). 
    
    \item \emph{Non-uniform active contraction with $\zeta_{\rm a}=\zeta_0 r/R$.} The exact analytical solution is
    \begin{equation}\label{Eq:App1-ThinSol2}
    \begin{aligned}
    u(r)&=-\frac{1}{18}\alpha{\cal Z}_{\rm b} \frac{h^2}{R^2}\left[(1-\nu_0)r+ (1+\nu_0) \frac{r^2}{R}\right], \\
    w(r)&=\frac{1}{6}\alpha(1-\alpha){\cal Z}_{\rm b}\frac{h}{R^2}\left[(1-\nu_0) r^2+\frac{2}{3}(1+\nu_0) \frac{r^3}{R}\right].
    \end{aligned}
    \end{equation}
    From Eq.~(\ref{Eq:App1-ThinSol2}), we obtain the two principal bending curvatures are not equal any more and take the dimensionless form of  $w''h=\frac{1}{3}\alpha(1-\alpha){\cal Z}_{\rm b}\frac{h^2}{R^2}\left[1-\nu_0 + 2(1+\nu_0) \frac{r}{R}\right]$ and $w'h/r=\frac{1}{3}\alpha(1-\alpha){\cal Z}_{\rm b}\frac{h^2}{R^2}\left[1-\nu_0 +(1+\nu_0) \frac{r}{R}\right]$, which are non-uniform through the circular plate and the sign depends on the position and Poisson's ratio as shown in Fig.~\ref{Fig:App1-Bending}(b).  
\end{itemize}

We now show that the above elastostatic problem of active circular plates can also be solved by Ritz's variational method of approximation~\cite{Reddy2017} as follows. Note that the trial functions we take have to satisfy the essential boundary conditions in the problem, {\emph{i.e.}}, $u(r=0)=w(r=0)=0$ and $w'(r=0)=0$.

\begin{itemize}
    \item \emph{Uniform active contraction with $\zeta_{\rm a}=\zeta_0$.} We assume (from physical intuitions) that, as a reasonable guess, the circular plate is contracted uniformly in-plane and bent with uniform curvature out-of-plane, that is, 
\begin{equation}\label{Eq:App1-RitzSol1}
u(r)=\epsilon_1 r, \quad
w(r)=\frac{1}{2} \kappa r^2,
\end{equation} 
in which $\epsilon_1$ and $\kappa$ are two constant parameters to be determined from energy minimization. Substituting them into Eq.~(\ref{Eq:App1-ThinFtuw}), we obtain the total energy as a function of $\epsilon_1$ and $\kappa$ as
\begin{equation}\label{Eq:App1-RitzFt1}
\begin{aligned}
{\cal F}_{\rm t}(\epsilon_1,\kappa) = 
&2\pi R^2 
\left[
\left(1+\nu_{0}\right)\left(\frac{Y}{2} \epsilon_1^2 + \frac{D}{2} \kappa^2 \right) -\alpha \zeta_0h\epsilon_1  \right. \\
&\left.+\frac{1}{2}\alpha(1-\alpha) \zeta_0  h^2\kappa 
\right]. 
\end{aligned}
\end{equation}

\begin{figure}
  \centering
  \includegraphics[width=0.75\linewidth]{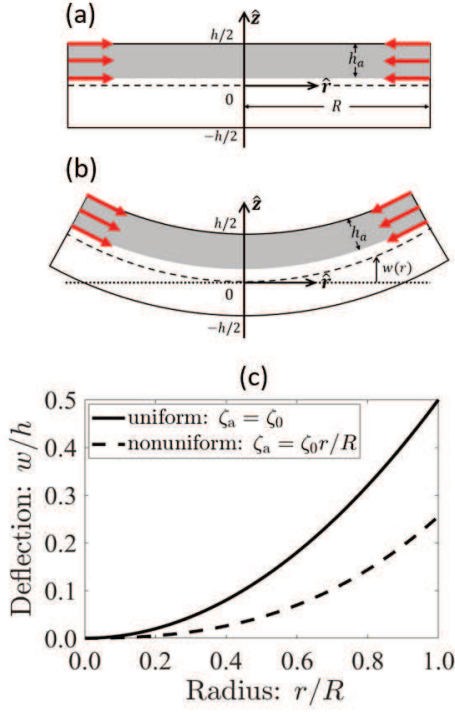}
  \caption {(color online) 
 (a, b) Spontaneous bending of a thin active circular plate of thickness $h$ and lateral radius $R$ from its reference state (a) to its deformed state (b). The plate is suspended by its center and the internal active contraction is distributed asymmetrically along the plate thickness. The thickness of the active contracting layer is $h_{\rm a}$ and its fraction is $\alpha=h_{\rm a}/h \in [0,1]$. (c) The deflection $w/h$ of the active plate as a function of radius $r/R$ for both uniform and non-uniform contraction. 
  }\label{Fig:App1-Bending}   
\end{figure}

Minimization of ${\cal F}_{\rm t}$ with respect to $\epsilon_1$ and $\kappa$ gives $\epsilon_1=-\frac{1}{6}\alpha{\cal Z}_{\rm b}\frac{h^2}{R^2}$ and the normalized curvature ${\cal K}=\kappa R^2/h=\alpha(1-\alpha) {\cal Z}_{\rm b}$. That is, we have used an alternative variational method to reproduce the exact solution as discussed near Eq.~(\ref{Eq:App1-ThinSol1}).  

\item \emph{Non-uniform active contraction with $\zeta_{\rm a}=\zeta_0 r/R$.} We follow Ritz's method to take two different forms of trial functions. Firstly, we choose the same trial function as Eq.~(\ref{Eq:App1-RitzSol1}) and then the total energy is given by
\begin{equation}\label{Eq:App1-RitzFt2}
\begin{aligned}
{\cal F}_{\rm t}(\epsilon_1,\kappa) =
& 2\pi R^2 \left[\left(1+\nu_{0}\right)\left(\frac{Y}{2} \epsilon_1^2 + \frac{D}{2} \kappa^2 \right)
-\frac{2}{3}\alpha \zeta_0h\epsilon_1  \right. \\
&\left. + \frac{1}{3}\alpha(1-\alpha) \zeta_0  h^2\kappa \right].  
\end{aligned}
\end{equation}

Minimization of ${\cal F}_{\rm t}$ with respect to $\epsilon_1$ and $\kappa$ gives $\epsilon_1=-\frac{1}{9}\alpha{\cal Z}_{\rm b}\frac{h^2}{R^2}$ and the normalized curvature ${\cal K}=\kappa R^2/h=2\alpha(1-\alpha){\cal Z}_{\rm b}/3$. 
Next, to get a better approximation, we try a trial function of higher order polynomials:
\begin{equation}\label{Eq:App1-RitzSol2}
u(r)=\epsilon_1 r+\frac{1}{2} \epsilon_2 \frac{r^2}{R}, \quad
w(r)=\frac{1}{2} \kappa_1 r^2+\frac{1}{3} \kappa_2 \frac{r^3}{R},
\end{equation}
from which we obtain ${\cal F}_{\rm t}$ as a function of four parameters, $\epsilon_1$, $\epsilon_2$, $\kappa_1$, and $\kappa_2$. Minimization of ${\cal F}_{\rm t}$ then reproduces the exact solution as given near Eq.~(\ref{Eq:App1-ThinSol2}) in the previous subsection.  
\end{itemize}

\subsubsection{Spontaneous bending at large deflections}\label{Sec:App1-Thin-Ritz}

We now consider the case of relatively large deflections compared with the plate thickness ($h$) but still small compared with the lateral radius $R$ of the plate, \emph{i.e.}, $h\ll w\ll R$. We still neglect the effects of gravitational forces with $\rho g=0$. In this case, we can't use Kirchhoff's plate theory but have to use von-k\'arm\'an's plate theory~\cite{Masters1994,Freund2000}: the nonlinear terms involving $w$ in the definitions of neutral-surface strain can not be neglected, while the nonlinear terms involving in-plane displacements are still assumed negligible. Hence, the displacement vector can still take the form of Eq.~(\ref{Eq:App1-Displacements}), but the neutral-surface strains are given by 
\begin{equation}\label{Eq:App1-DeepRitzStrains}
\epsilon_{\rm rr}=u'+\frac{1}{2}w'^2-zw'', \quad
\epsilon_{\rm \theta\theta}=u/r-zw'/r.
\end{equation}
which include nonlinear term $\frac{1}{2}w'^2$ and are different from Eq.~(\ref{Eq:App1-ThinStrains}) for small deflections. Substituting them into Eq.~(\ref{Eq:App1-ThinFt}) and integrating over thickness $z$-direction, we obtain the total energy as a functional of $u(r)$ and $w(r)$: 
\begin{equation}\label{Eq:App2-ThinFtuw}
\begin{aligned}
&{\cal F}_{\rm t}[u(r),w(r)]= \\
& \int_0^R dr 2\pi r \left\{\frac{Y}{2}\left[\left(u'+\frac{1}{2}w'^2 \right)^2  +\frac{u^2}{r^2}+2\nu_0 \frac{u}{r} \left(u'+\frac{1}{2}w'^2 \right)\right] \right.\\
& - \tau_{\rm a}\left(u'+\frac{1}{2}w'^2+\frac{u}{r}\right)
+\frac{D}{2}\left(w''^2 +\frac{w'^2}{r^2}+2\nu_0 \frac{w'w''}{r}\right)\\
&\left. -M_{\rm a}\left(w''+\frac{w'}{r}\right) \right\}.
\end{aligned}
\end{equation}

Here we only consider uniform active contraction with $\zeta_{\rm a}=\zeta_0<0$. Using Ritz method, we take the trial function~\cite{Masters1994,Freund2000} as
\begin{equation}\label{Eq:App1-DeepRitzSol1}
u(r)=\epsilon_1 r+\epsilon_2 \frac{r^3}{R^2}, \quad
w(r)=\frac{1}{2} \kappa r^2,
\end{equation} 
in which $\epsilon_1$, $\epsilon_2$, and $\kappa$ are three undetermined  parameters. Substituting them into Eq.~(\ref{Eq:App2-ThinFtuw}), we obtain the total energy as a function of the three parameters as
\begin{equation}\label{Eq:App1-DeepRitzFt1}
\begin{aligned}
&{\cal F}_{\rm t}(\epsilon_1,\epsilon_2,\kappa) = 2\pi R^2 \left[(1+\nu_{0})\frac{Y}{2} \left(\epsilon_1^2+2\epsilon_1\epsilon_2+\frac{5+3\nu_0}{3+3\nu_0}\epsilon_2^2\right) \right. \\
& + \frac{D}{2}(1+\nu_0) \left(1+3 \epsilon_1 \frac{R^2}{h^2} + 2\epsilon_2\frac{3+\nu_0}{1+\nu_0} \frac{R^2}{h^2} \right) \kappa^2 + \frac{D}{4h^2}\kappa^4 R^4 \\
& \left. -\alpha \zeta_0h\left(\epsilon_1 +\epsilon_2-\frac{1}{2}(1-\alpha)\kappa h+\frac{1}{8}\kappa^2R^2 \right)
\right].
\end{aligned}
\end{equation}

Minimization of ${\cal F}_{\rm t}$ with respect to $\epsilon_1$, $\epsilon_2$, and $\kappa$ gives 
\begin{equation}\label{Eq:App1-DeepRitzPara}
\begin{aligned}
\epsilon_1 =-\frac{1}{6}\alpha{\cal Z}_{\rm b} \frac{h^2}{R^2} 
&+\frac{1-\nu_0}{16}\frac{h^2}{R^2}{\cal K}^2, \qquad
\epsilon_2 =-\frac{3-\nu_0}{16}\frac{h^2}{R^2}{\cal K}^2, \\
& {\cal K}+\frac{1-\nu_0}{16}{\cal K}^3-\alpha(1-\alpha){\cal Z}_{\rm b}=0. 
\end{aligned}
\end{equation}
with ${\cal Z}_{\rm b}$ defined in Eq.~(\ref{Eq:App1-Zcb}) and the normalized curvature ${\cal K}=\kappa R^2/h$.

\subsection{Deep Ritz method (DRM)}\label{Sec:App1-DeepRitz}

In this subsection, we use the deep Ritz method to study the spontaneous bending of thin active circular plates. Firstly for small deflections, we de-dimensionalize the free energy in Eq.~(\ref{Eq:App1-ThinFtuw}) by taking $\tilde{r}=r/R$, $\tilde{u}=u R/h^2$. The dimensionless total free energy is then given by
\begin{equation}\label{Eq:App1-ThinFtuw-ML1}
\begin{aligned}
&\frac{{\cal F}_{\rm t}[\tilde{u}(\tilde{r}),\tilde{w}(\tilde{r})]}{E_0\pi h^5/R^2(1-\nu_0^2)} 
= \int_0^1 d\tilde{r} \tilde{r} \left[\left(\tilde{u}'^2 +\tilde{u}^2/\tilde{r}^2+2\nu_0 \tilde{u}\tilde{u}'/\tilde{r}\right) \right.\\
& +\frac{1}{3}\alpha{\cal Z}_{\rm b} (1+\nu_0)(\tilde{u}'+\tilde{u}/\tilde{r})+ \frac{1}{12}\left(\tilde{w}''^2 +\tilde{w}'^2/\tilde{r}^2+2\nu_0 \tilde{w}'\tilde{w}''/\tilde{r}\right) \\
&\left. -\frac{1}{6}\alpha(1-\alpha){\cal Z}_{\rm  b} (1+\nu_0)(\tilde{w}''+\tilde{w}'/\tilde{r})\right]
\end{aligned}
\end{equation}
with $\tilde{u}'=d\tilde{u}/d\tilde{r}$, $\tilde{w}=w/h$, $\tilde{w}'=d\tilde{w}/d\tilde{r}$. We would like to point out that proper de-dimensionalization is important technically for fast convergence in DRM. 

Following the DRM explained in Sec.~\ref{Sec:VarMeth-DeepRitz}, we first check the performance of the neural network with different parameter settings. As shown in Table~\ref{table:eg1}, one can observe that the $L_\infty$ error and the mean square errors away from the exact solution are all very small when width $m$, depth $N_{\rm n}$, and $N_s$ are relatively large. Particularly, the small $L_\infty$ error about $10^{-5}$ implies a high accuracy of the predicted solution from the neural network. Moreover, we note that the errors are not exactly monotonically decreasing but oscillating in a very small region as $m$, $N_{\rm n}$, or $N_s$ increases, which is reasonable because of the random generation of spatial (input observation) points and the application of stochastic gradient descent algorithms. Furthermore,  we find that for a given setting of the network, there seems to exist a minimal training steps (or training time, say $T_{\rm s}$) to obtain a relatively small and stable training error (\emph{i.e.}, the value of the error oscillates around a very small value). 
Importantly, we find that as $m$, $N_{\rm n}$, or $N_s$ increase, the minimal training time $T_{\rm s}$ seems to increase linearly or even sublinearly, and the training results will not get ``worse'' (for example, no larger errors or over-fitting) as long as the training iterates over the corresponding minimal $T_{\rm s}$ steps.  

\begin{figure}[ht!]
  \centering
\subfigure{\includegraphics[width = 0.7 \linewidth]{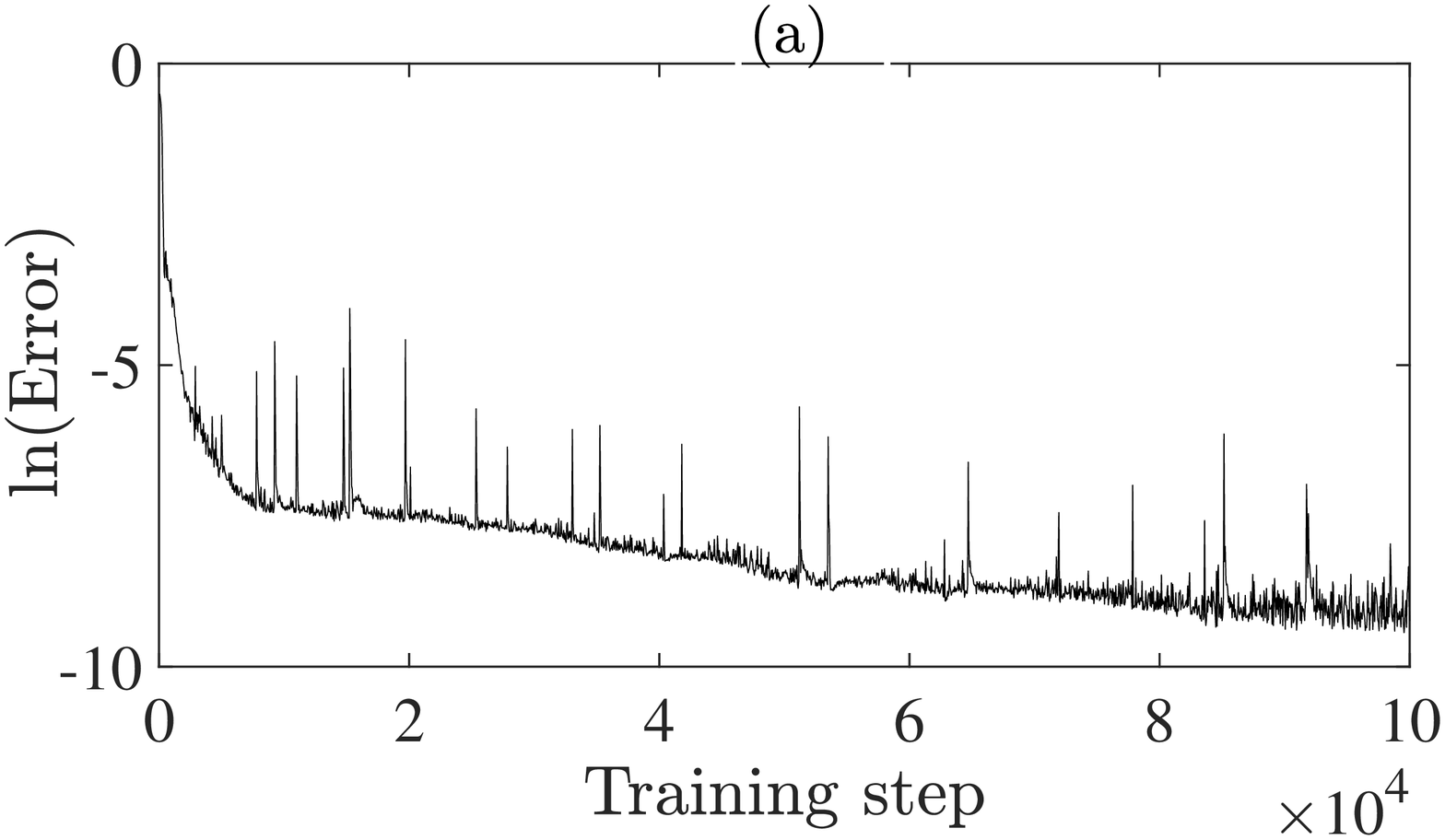}}
\subfigure{\includegraphics[width = 0.7 \linewidth]{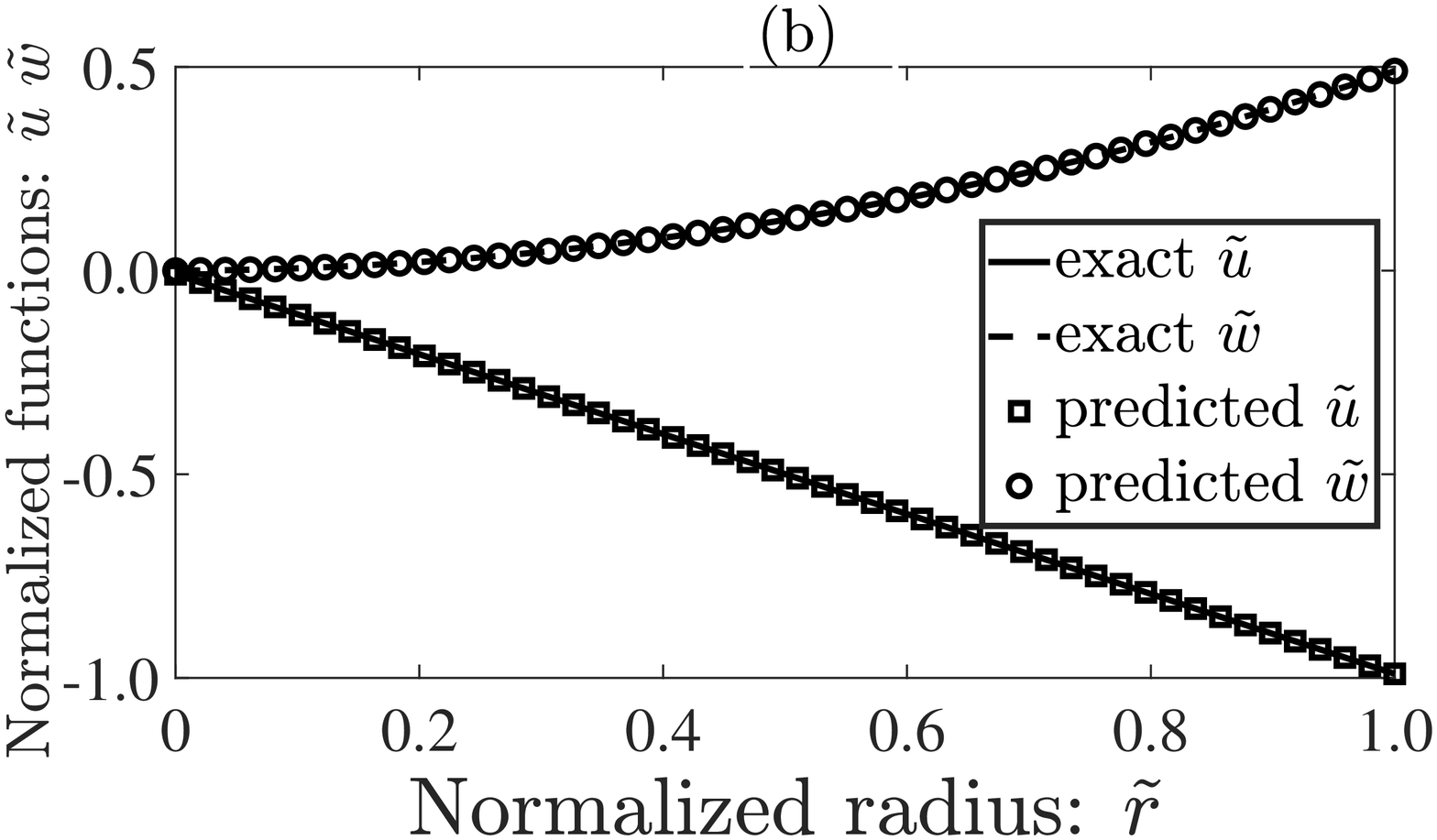}}
\subfigure{\includegraphics[width = 0.7 \linewidth]{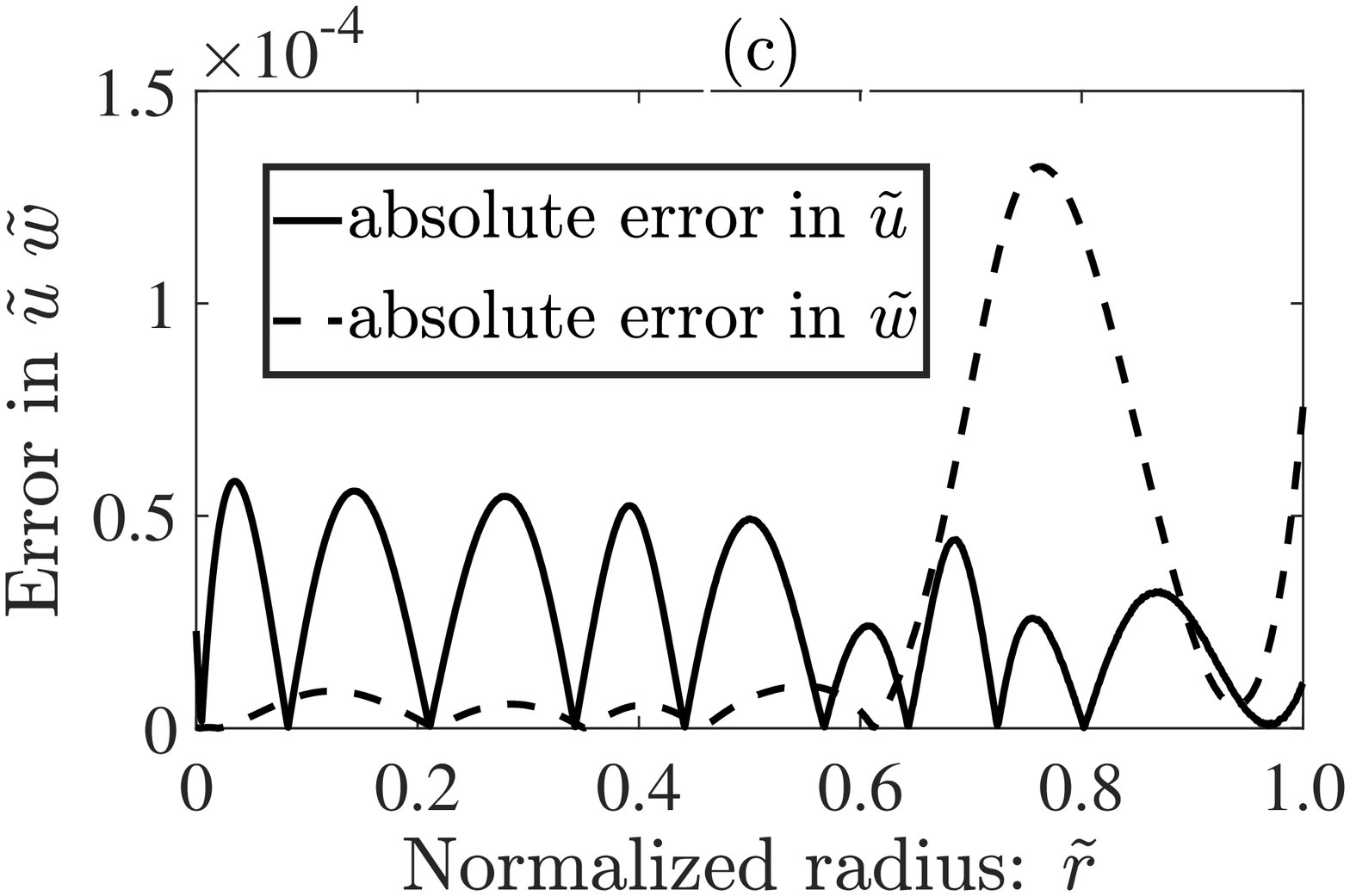}}
  \caption {Numerical results predicted using the deep Ritz method for the bending of active circular plate at small deflections. (a) The change of ${\rm Error}$ (as defined in Table~\ref{table:eg1}) with respect to $\tilde{u}$ and $\tilde{w}$ during the training process at every 50 training steps. (b) The predicted results and their good agreement with the exact solutions in Eq.~(\ref{Eq:App1-ThinSol1}). (c) The absolute error between the final predicted solution and the exact solution at the corresponding $\tilde{r}$ in Eq.~(\ref{Eq:App1-ThinSol1}). Here we take $\alpha=5/6$ and $\alpha(1-\alpha){\cal Z}_{\rm b}=1$.
  }\label{Fig:App1-ML1}   
\end{figure}

In addition, we have also varied the activation functions, for example, we have tried ReLU, ReLU$^2$, ReLU$^3$, Tanh, Tanh$^3$, and Sigmoid functions. We found that ReLU$^3$ and Tanh$^3$ outperform others and hence we use ReLU$^3$ to carry out all the training studies as shown in this section. 
In Fig.~\ref{Fig:App1-ML1}, we plotted the temporal evolution of training errors during the training process and the predicted results from the neural network after $100, 000$ training steps with $m$, $N_{\rm n}$, $N_{s}$ chosen to be $10$, $4$, $1000$, respectively, and $l_{\rm r}=0.001$ (decreased to be half at some milestone training steps). The predicted results agree very well with the exact solution in Eq.~(\ref{Eq:App1-ThinSol1}) as shown in  Fig.~\ref{Fig:App1-ML1}(b). 

\begin{table}[ht]
\large
\begin{tabular}{c|c|c|c|c|c|c}
m  & $N_{\rm n}$ & $N_{s}$   & $N$    & $E^{\tilde u}_\infty$      & $E^{\tilde w}_\infty$      & $\rm Error$      \\ \hline
5  & 4  & 1000 & 132   & 0.000074 & 0.000448 & 0.000241 \\
10 & 4  & 1000 & 462   & 0.000058 & 0.000264 & 0.000097 \\
20 & 4  & 1000 & 1722  & 0.000019 & 0.000117 & 0.000068 \\
40 & 4  & 1000 & 6642  & 0.000056 & 0.000156 & 0.000076 \\
80 & 4  & 1000 & 26082 & 0.000032 & 0.000188 & 0.000082 \\
\hline
10 & 1  & 1000 & 132   & 0.000379 & 0.036926 & 0.024460 \\
10 & 2  & 1000 & 242   & 0.000049 & 0.000162 & 0.000091 \\
10 & 4  & 1000 & 462   & 0.000058 & 0.000264 & 0.000097 \\
10 & 6  & 1000 & 682   & 0.000064 & 0.000245 & 0.000107 \\
10 & 10 & 1000 & 1122  & 0.000109 & 0.000173 & 0.000082 \\
\hline
10 & 4  & 10   & 462   & 0.000569 & 0.001046 & 0.000547 \\
10 & 4  & 100  & 462   & 0.000093 & 0.000424 & 0.000181 \\
10 & 4  & 500  & 462   & 0.000089 & 0.000159 & 0.000100 \\
10 & 4  & 1000 & 462   & 0.000058 & 0.000264 & 0.000097 \\
10 & 4  & 5000 & 462   & 0.000066 & 0.000384 & 0.000156
\end{tabular}
\caption{Performance check of the neural network with different parameter settings: $m$ is network width, $N_{n}$ is network depth, $N_{s}$ is the number of input spatial coordinate points used at every training step, and $N$ is the total number of undetermined parameters. The $L_\infty$ error and the mean square errors of the training predictions away from exact solution are computed. $E^{\tilde u}_\infty=max(abs(\tilde{u}-\tilde{u}_{exact}))$ is the $L_\infty$ error in $\tilde u$ computed on $1000$ uniform points in $[0,1]$, similarly $E^{\tilde w}_\infty=max(abs(\tilde{w}-\tilde{w}_{exact}))$, and ${\rm Error}=MSE_{\tilde u} + MSE_{\tilde w}$ is the summation of the mean square error in $\tilde u$ and $\tilde w$. Here we set the training step to be $50, 000$ and the learning rate $l_{\rm r}=0.001$ (decreased to be half at some milestone training steps).}
\label{table:eg1} 
\end{table}

For relatively large deflections with $h\ll w\ll R$, we use the same dimensionless methods and the dimensionless form of the total free energy in Eq.~(\ref{Eq:App2-ThinFtuw}) is then given by 
\begin{equation}\label{Eq:App1-ThinFtuw-ML2}
\small{
\begin{aligned}
&\frac{{\cal F}_{\rm t}[\tilde{u}(\tilde{r}),\tilde{w}(\tilde{r})]}{E_0\pi h^5/R^2(1-\nu_0^2)}\\
&=\int_0^1 d\tilde{r} \tilde{r} \left\{\left[\left(\tilde{u}'+\frac{1}{2}\tilde{w}'^2 \right)^2  +\frac{\tilde{u}^2}{\tilde{r}^2} +2\nu_0 \frac{\tilde{u}}{\tilde{r}} \left(\tilde{u}'+\frac{1}{2}\tilde{w}'^2 \right)\right]\right.\\
& + \frac{1}{3}\alpha{\cal Z}_{\rm b} (1+\nu_0)\left(\tilde{u}'+\frac{1}{2}\tilde{w}'^2+\frac{\tilde{u}}{\tilde{r}}\right) + \frac{1}{12}\left(\tilde{w}''^2 +\frac{\tilde{w}'^2}{\tilde{r}^2}+2\nu_0 \frac{\tilde{w}'\tilde{w}''}{\tilde{r}}\right) \\
&\left.-\frac{1}{6}\alpha(1-\alpha){\cal Z}_{\rm  b} (1+\nu_0)\left(\tilde{w}''+\frac{\tilde{w}'}{\tilde{r}}\right) \right\}.
\end{aligned}}
\end{equation}

In this case, the deformation of the active circular plate is more complicated and the exact solution is not available. But in Sec.~\ref{Sec:App1-Thin-Ritz}, we have used the classical Ritz method to derive the first-order approximate solution as those in Eq.~(\ref{Eq:App1-DeepRitzPara}) by taking the simple polynomial trial function in Eq.~(\ref{Eq:App1-DeepRitzSol1}). 

\begin{figure}[htbp]
  \centering
\subfigure{\includegraphics[width = 0.75 \linewidth]{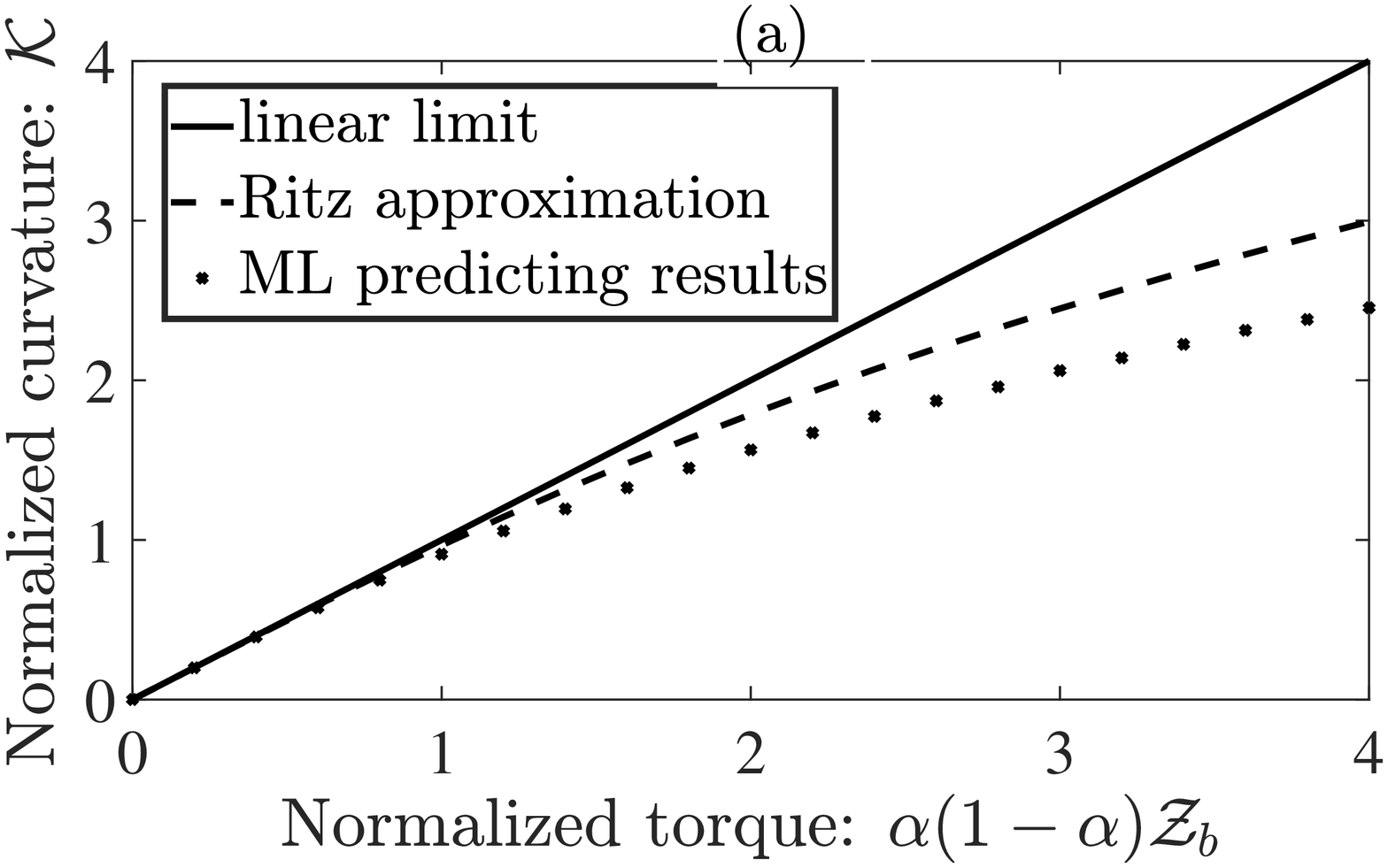}}
\subfigure{\includegraphics[width = 0.75 \linewidth]{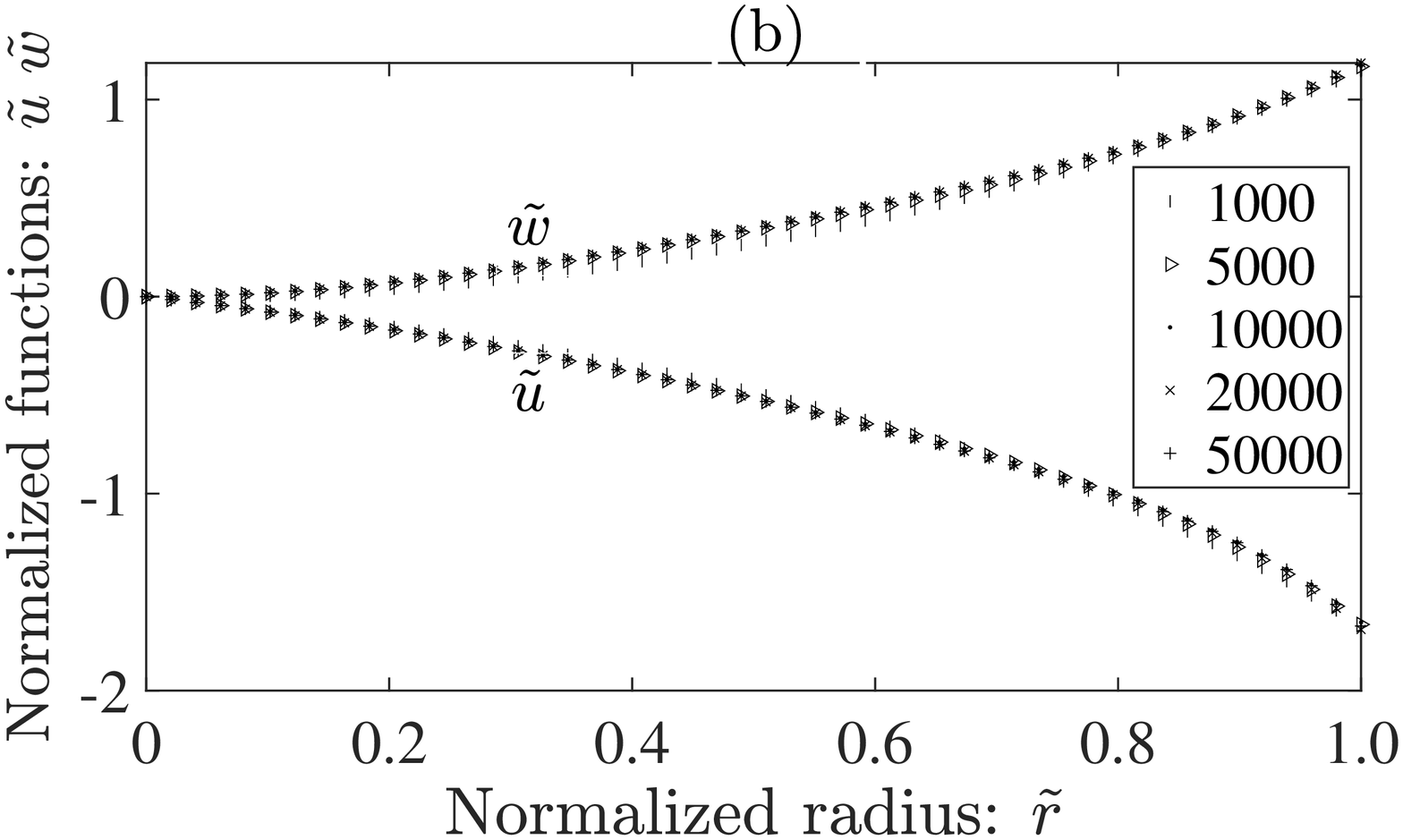}}
\subfigure{\includegraphics[width = 0.75 \linewidth]{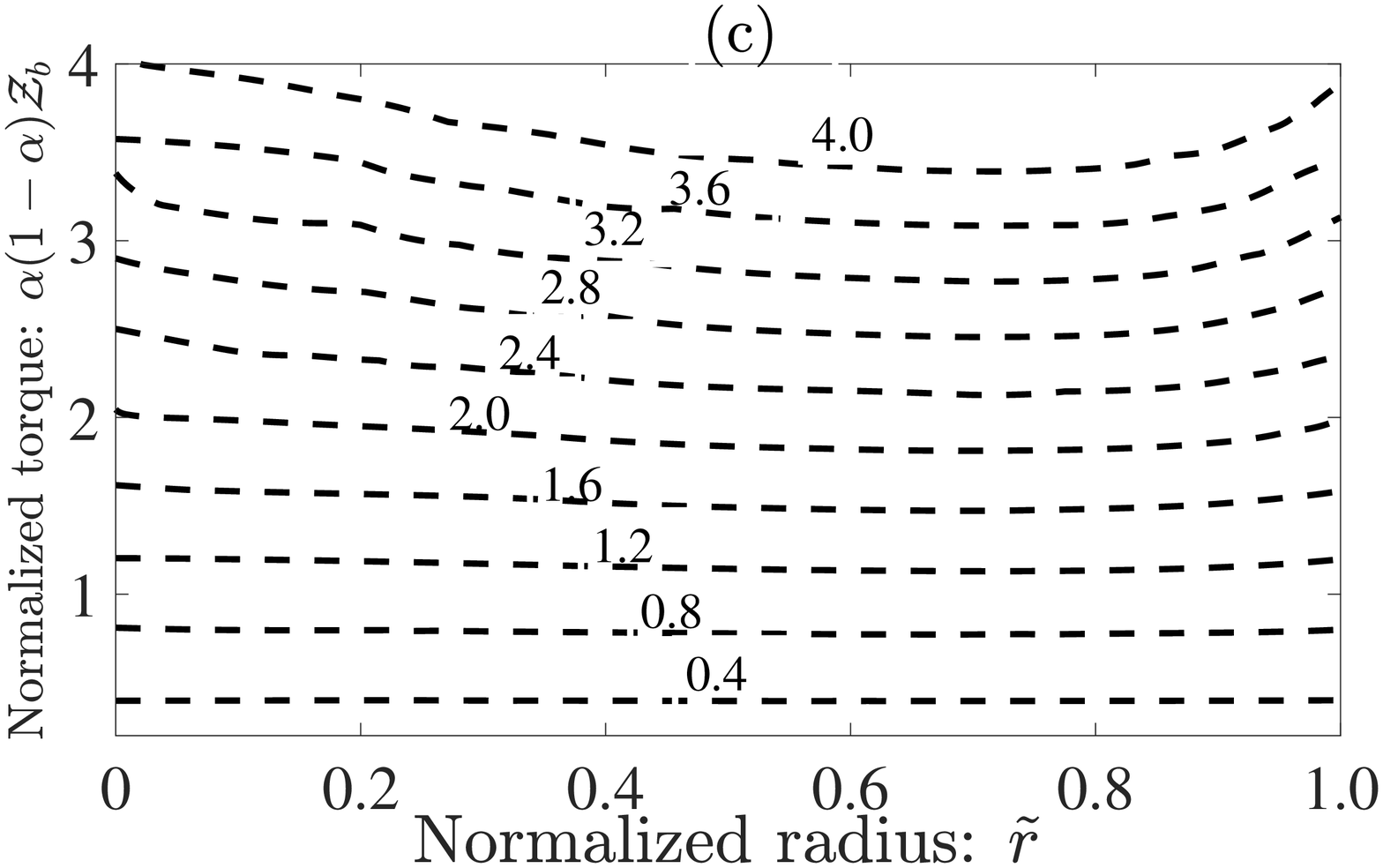}}
\caption{Numerical results predicted using the DRM for the bending of active circular plate at large deflections. (a) The change of normalized curvature ${\cal K}=\kappa R^2/h$ with normalized torque $\alpha(1-\alpha){\cal Z}_{\rm b}=-\alpha(1-\alpha){\zeta_0 h R^2}/{2D(1+\nu_0)}$ for large deflections with spatially uniform substrate curvature. The solid line represents the linear relation implied by small deflection theory in Eq.~(\ref{Eq:App1-ThinSol1}). The dashed line represents Ritz's approximate solution in Eq.~(\ref{Eq:App1-DeepRitzFt1}). (b) Predicted solutions obtained at several different training steps: 1,000, 5,000, 10,000, 20,000, and 50,000, with $\alpha=1/3$ and  $\alpha(1-\alpha){\cal Z}_{\rm b}=4$. (c) The contour plot of normalized curvature ${\cal K}$ in the plane of normalized radius and normalized torque. A spatially uniform substrate curvature corresponds to a horizontal contour line. Here we take $\alpha=1/3$.}
\label{Fig:App1-ML2}
\end{figure}

Here we use the deep Ritz method to numerically solve the variational problem by using the same neural network structure as above, in which the network parameters are now optimized by minimizing the total free energy in Eq.~(\ref{Eq:App1-ThinFtuw-ML2}) as the loss function for several different values of ${\cal Z}_{\rm b}=-{\zeta_0 h R^2}/{2D(1+\nu_0)}$ defined in Eq.~(\ref{Eq:App1-Zcb}). In Fig.~\ref{Fig:App1-ML2}, we plot the normalized curvature ${\cal K}=\kappa R^2/h$ as a function of the normalized torque $\alpha(1-\alpha)\cal Z_{\rm b}$ and its contours in the plane of normalized radius $\tilde{r}=r/R$ and
normalized torque $\alpha(1-\alpha)\cal Z_{\rm b}$. In Fig.~\ref{Fig:App1-ML2}(a), the predicted normalized curvature is obtained from a least-square fitting of the predict solution $w(r)$ by $\frac{1}{2}\kappa r^2$. We can see that when $\cal Z_{\rm b}$ is relatively small, the solution agrees well with the Ritz approximation because the exact solution would be closer to a quadratic function as assumed in Ritz approximation in Eq.~(\ref{Eq:App1-DeepRitzSol1}). For larger $\cal Z_{\rm b}$, the predict solution from the deep Ritz method becomes more complicated and deviates from Ritz's approximate solution in Eq.~(\ref{Eq:App1-DeepRitzSol1}).  In Fig.~\ref{Fig:App1-ML2}(b), we plot the predicted solutions at several different training steps, implying the convergence of the predicted solution. Indeed, one can observe in Fig.~\ref{Fig:App1-ML2}(c) that, for a large fixed $\cal Z_{\rm b}$, the normalized curvature is not a constant anymore but changes with the normalized radius. The self-consistency of the predicted results justifies the ability of deep Ritz method to explore high dimensional information or more complicated behaviors. However, the accuracy of these results should be further checked and compared quantitatively with finite element simulations or a direct numerical solutions of the equilibrium equations at large deflections. We leave this to our future work. 

\subsection{Gravitaxis of thin active circular plates}\label{Sec:App1-Gravi}

Gravitaxis is a form of cell taxis characterized by the directional movement of cells in response to gravitational forces~\cite{Schwartzbach2017}. There are many different biological and physical mechanisms causing the gravitaxis of individual cells. For example, some cells have receptors like statocysts that allow them to sense the gravitational force and to adjust their body orientation or migrating directions accordingly~\cite{Schwartzbach2017,Bechinger2014}. Some other cells do not have gravisensory structures, in which case gravitaxis can result from a purely physical mechanism such as asymmetric mass distribution in the cell body.  
Here, however, we consider non-motile adherent animal cells and propose theoretically a possible way of sensing gravitational forces in cell aggregates or tissue scales. For this purpose, we study a toy system: a thin active circular plate with supported edges placed horizontally in a gravitational field. The active circular plate can be regarded as a minimal model of cell monolayer. 


\begin{figure}
  \centering
  \includegraphics[width=0.35\textwidth]{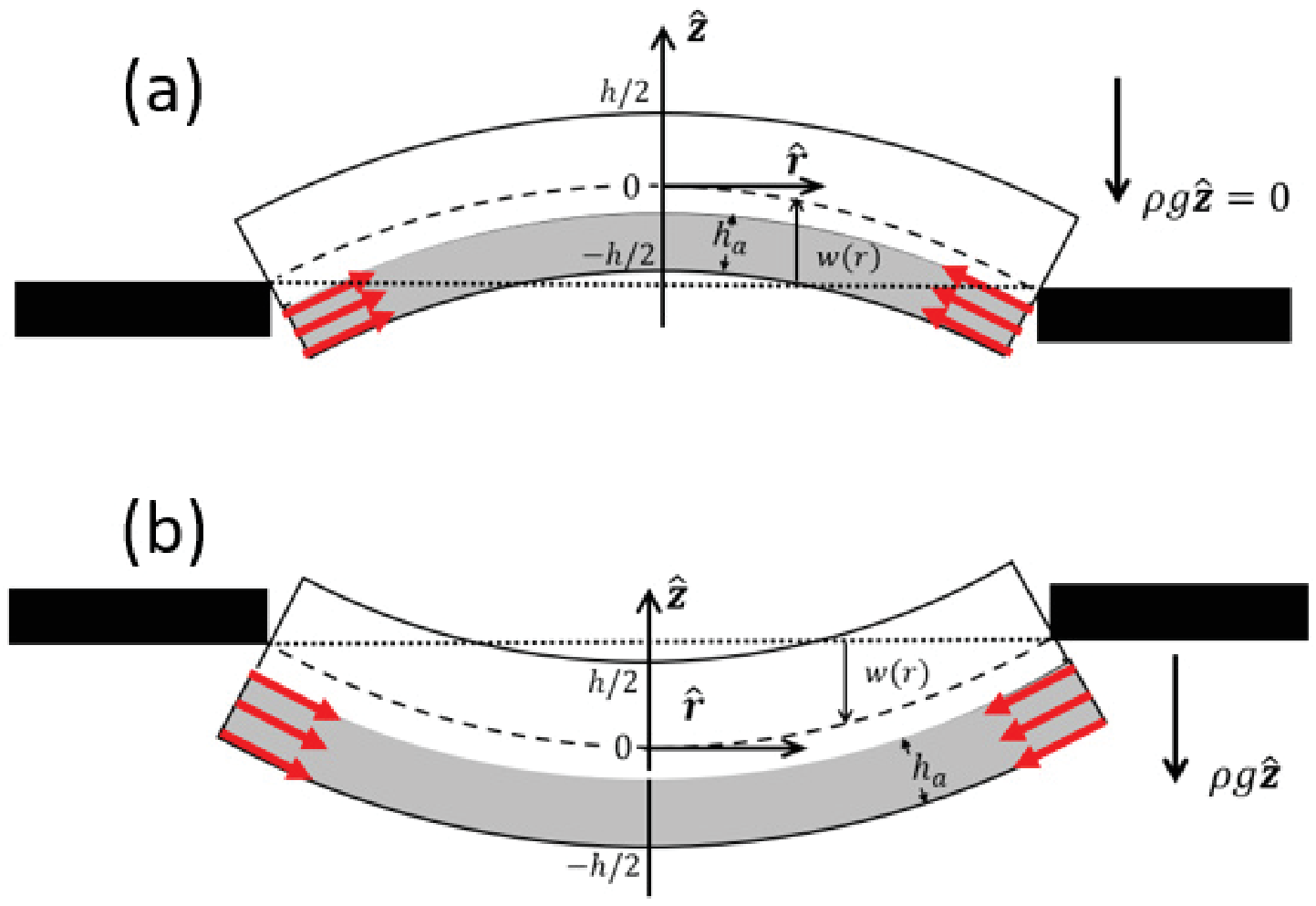} 
  \includegraphics[width=0.35\textwidth]{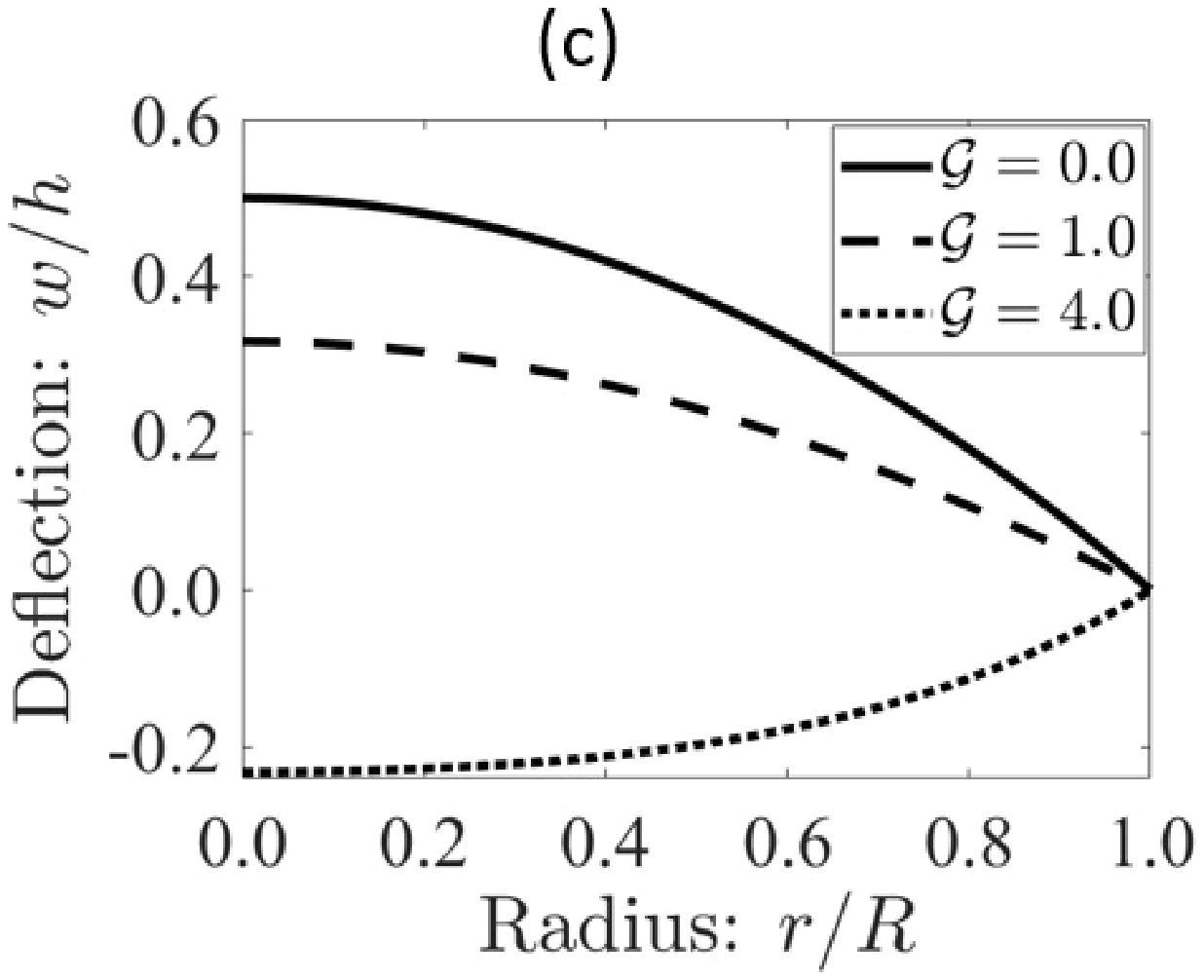} 
    \includegraphics[width=0.33\textwidth]{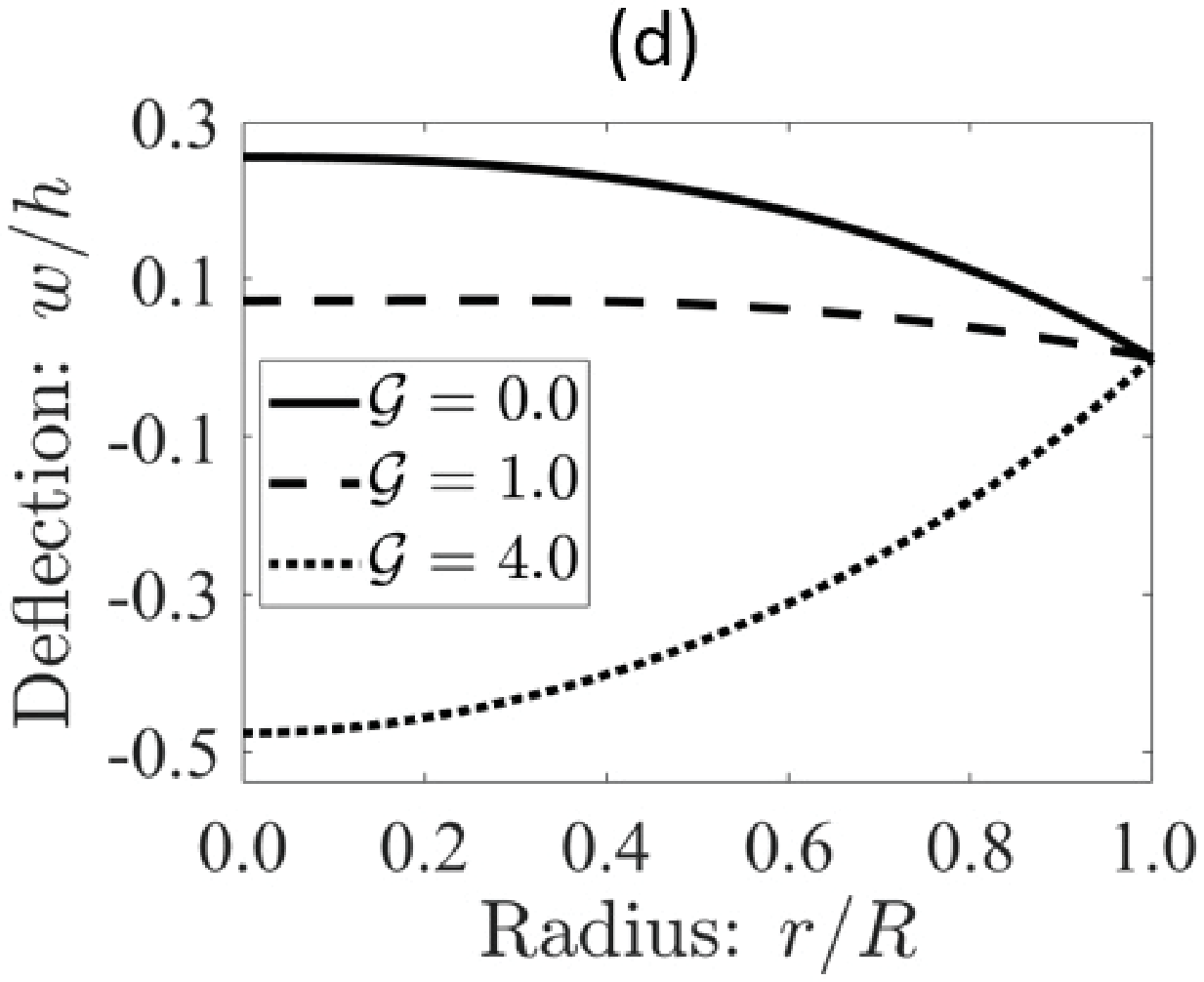} 
  \caption {(color online) 
Spontaneous bending of thin active circular plates with supported edges at (a) zero gravity and (b) non-zero gravity. In the case of non-zero gravity, the deflections $w/h$ of the bending plate are calculated as a function of $r/R$ under different activogravity number $\cal G$ for both (c) uniform and (d) non-uniform active contraction. When ${\cal G}> 1$, gravitational forces are non-negligible in bending the plate. 
  }\label{Fig:App1-Gravi}  
\end{figure}

In order to better show the competition between active cellular forces and gravitational forces, we consider the circular plate shown in Fig.~\ref{Fig:App1-Gravi} where the internal active contraction tends to bend the plate upward (see Fig.~\ref{Fig:App1-Gravi}(a)), while the gravitational force tends to bend it downward (see Fig.~\ref{Fig:App1-Gravi}(b)). In this case, the active stress component is given (in a slightly different form from Eq.~(\ref{Eq:App1-Zeta})) by
\begin{align}\label{Eq:App3-Zeta}
\zeta = \zeta_{\rm a} \Theta(-z-(1-2 \alpha)\frac{h}{2}),  
\end{align}
with $\zeta_{\rm a}=\zeta_0<0$ and $\zeta_{\rm a}=\zeta_0 r/R<0$ for uniform and non-uniform contraction, respectively. Furthermore, to focus on the physical mechanisms, we only consider small deflections with $w<<h$, in which the components of displacement vector and strain tensor are given by Eqs.~(\ref{Eq:App1-Displacements}) and (\ref{Eq:App1-ThinStrains}). Substituting them into Eq.~(\ref{Eq:App1-ThinFt}) for $\rho g\neq 0$ and integrating over thickness $z$-direction, we obtain the total energy functional:
\begin{equation}\label{Eq:App3-ThinFtuw}
\begin{aligned}
&{\cal F}_{\rm t}[u(r),w(r)] 
= \int_0^R dr 2\pi r 
\left[\frac{Y}{2} \left(u'^2 +u^2/r^2+2\nu_0 uu'/r\right)\right. \\
& - \tau_{\rm a} (u'+u/r)
+ \frac{D}{2}\left(w''^2 +w'^2/r^2+2\nu_0 w'w''/r\right) 
\left.-M_{\rm a} (w''+w'/r)+\rho g w\right].
\end{aligned}
\end{equation}
Then from the minimization of ${\cal F}_{\rm t}$ with respect to $u(r)$ and $w(r)$ gives the equilibrium equations (\emph{i.e.}, the Euler-Lagrange equations) in the bulk (for $0\leq r\leq R$) as Eq.~(\ref{Eq:App1-ThinEqns1}) and
\begin{subequations}\label{Eq:App3-ThinEqns12345}
\begin{equation}\label{Eq:App3-ThinEqns2}
D\left( w^{(4)}+\frac{2w'''}{r}-\frac{w''}{r^2}+\frac{w'}{r^3}\right)
+\left( M''_{\rm a}+\frac{M'_{\rm a}}{r} \right)=-\rho gh,
\end{equation} 
and boundary conditions as in Eq.~(\ref{Eq:App1-ThinEqns3}) and 
\begin{equation}\label{Eq:App1-GraviBcs}
w|_{r=R}=0, \quad \left[D(w''+\nu_0\frac{w'}{r})+ M_{\rm a}\right]_{r=R}=0,
\end{equation} 
\end{subequations}
at the supported edges at $r=R$, which are supplemented with conditions at $r=0$: $u=0$, $w={\rm finite}$, and $w'=0$.
Particularly for both uniform and non-uniform active stresses discussed in Eq.~(\ref{Eq:App1-Zeta}), the above equation system can be both solved analytically as follows.

\emph{(1) Uniform active contraction with $\zeta_{\rm a}=\zeta_0<0$.} The exact analytical solution is
\begin{equation}\label{Eq:App1-GraviSol1}
\begin{aligned}
w(r)
&=\frac{{\cal Z}_{\rm b}h }{2}\left[\frac{{\cal G}(1+\nu_0)}{16} \left(1-\frac{r^4}{R^4}\right) \right.\\
& \left. +\left(\alpha (1-\alpha)-\frac{3+\nu_0}{8}{\cal G}\right) \left(1-\frac{r^2}{R^2}\right)\right],
\end{aligned}
\end{equation} 

\emph{(2) Non-uniform active contraction with $\zeta_{\rm a}=\zeta_0 r/R<0$.} The exact analytical solution is 
\begin{equation}\label{Eq:App1-GraviSol2}
\begin{aligned}
&w(r)=\frac{{\cal Z}_{\rm b}h }{2}
\left[\frac{{\cal G}(1+\nu_0)}{16}\left(1-\frac{r^4}{R^4}\right)+ \frac{2}{9}\alpha (1-\alpha)(1+\nu_0) \right. \\
&\left(1-\frac{r^3}{R^3}\right) 
\left.+ \left(\frac{1}{3}\alpha (1-\alpha)(1-\nu_0)-\frac{3+\nu_0}{8}{\cal G}\right) \left(1-\frac{r^2}{R^2}\right)\right].
\end{aligned}
\end{equation}  
Here the dimensionless parameter
\begin{equation}\label{Eq:App1-GraviNo}
{\cal G} \equiv  \frac{\rho g hR^2}{|\zeta_0| h^2} = \left( \frac{R}{R_{\rm g}} \right)^2,
\end{equation}
is denoted as \emph{activo-gravity number}, which measures the strength of gravitational force in bending the plate relative to the asymmetric active contractile forces inside the cell. An \emph{activo-gravity length} $R_{\rm g} \equiv (|\zeta_0| h/\rho g)^{1/2}$ is introduced, which tells that the bending of the plate can be significantly varied by the gravitational force when its lateral dimension $R$ is larger than $R_{\rm g}$, or equivalently ${\cal G} >1$ as shown in Fig.~\ref{Fig:App1-Gravi}. If we assume the active stress $|\zeta_0|\sim 1 {\rm kPa}$ and take $\rho\sim 1.0 {kg/m^3}$, $g\sim 10 \, {\rm m/s^2}$, and $h\sim 100 \, {\rm nm}$, then we get $R_{\rm g} \sim 100 \, {\rm \mu m}$, about several cell diameters. Biologically, this indicates that gravitaxis may be negligible at the length scale of individual cells, but gravitaxis becomes non-negligible and important at larger length scale, for example, during tissue development the gravitational forces may play important roles in guiding the tissue morphogenesis and in breaking the up-down symmetry.

In addition, it is interesting to note that the activogravity length takes similar form as the capillary length $\lambda_{\rm c} = (\gamma/\rho g)^{1/2}$ with $\gamma$ being the surface tension and the activogravity number ${\cal G}$ takes similar form as Bond number or E\"{o}tv\"{o}s number that represents the ratio between the buoyancy forces and surface tension of the liquid.

\section{Conclusion and remarks}\label{Sec:Conclude}

Variational methods have been widely used in the physical modeling of soft matter. These methods are mostly based on two variational principles of statistical thermodynamics: the variational principle of minimum free energy (MFEVP) for static problems\cite{Doi2013,Sam2018} and Onsager's variational principle (OVP) for dynamic problems~\cite{Onsager1931a,Onsager1953,Doi2013,Doi2020,Xu2021,Komura2022}. In our former work~\cite{Xu2021}, we have explored the variational methods that are based on OVP for the modeling of active matter dynamics. Here in this work, we have focused on the variational methods that are based on MFEVP and can be used for the modeling of static problems in active solids.  

Active solids, consisting of elastically coupled active agents, combine the central properties of passive elastic solids and active fluids. We showed that MFEVP can be used, on the one hand, to derive thermodynamically consistent continuum models for active solids, including both equilibrium equations and matching boundary conditions. On the other hand, direct variational methods such as the Ritz method can be developed to approximate the state of mechanical equilibrium where active stresses are balanced by elastic stresses. Interestingly, we showed that the idea of Ritz method for active solids can be further powered by the deep learning method if we use deep neural networks to construct the trial functions of the variational problems. 
These variational methods are then applied to study the spontaneous bending and contraction of active circular plates, which can be regarded as a minimal model of a solid-like confluent cell monolayer. We found that circular plates can be bent by asymmetric active contraction inside the plate. We proposed that the importance of gravitational force relative to active cellular force is characterized by an activogravity length, $R_{\rm g}$ (about 100 micron), analogous to capillary length. When the lateral dimension of the plate is larger than $R_{\rm g}$, gravitational forces become important in bending the plate. We therefore propose that such gravitaxis may play significant roles in the morphogenesis and in breaking the up-down symmetry during tissue development.
 
Below we make a few general remarks and outlook.

(i) \emph{Applications to tissue morphogenesis dynamics.} The calculations of single-layer circular plates give some implications on the spontaneous bending and contraction of cell monolayer. These calculations can be further extended to study the dynamics of tissue morphogenesis by considering dissipative processes and multilayered composite plates that mimic tissue structure~\cite{Ackermann2022}. 

(ii) \emph{Taking into account of the elasto-active feedback.} In this work, we have only considered active stresses with frozen, given distributions. In biology, the system activity usually couples closely with mechanics. For example, the orientation of cytoskeleton and the density of cells also change with local deformation, forces, and torques. Cell activity shows mechanochemical couplings with biochemical molecules such as morphogens~\cite{Kinjal2016}. Cells divisions and apoptosis also induce active forces and changes in space and time~\cite{Joanny2010}. Tensional homeostatic behaviors are observed at both single cell and tissue levels~\cite{Eastwood1998}. These new physics can be included either by considering nonlinear elasticity~\cite{Xu2015PRE,Yair2021}, biological penalty~\cite{Sam2013a}, or by introducing more slow variables~\cite{Joanny2015,Xu2021}, such as polarization or nematic order, and chemical compositions. More physical couplings can be incorporated either in the free energy (for reversible or energetic couplings) by using MFEVP or in dissipation functions (for irreversible or dissipative couplings) by using OVP. 

(iii) \emph{Deep learning-based numerical methods for solving variational problems in active matter physics.} The deep Ritz method mentioned above for the statics of active matter is developed~\cite{E2018} by combining Ritz's variational method of approximation~\cite{Reddy2017} with deep learning methods that are based on deep neural networks and stochastic gradient descent algorithms. Similar deep learning methods can also be developed~\cite{Jeff2018,Jeff2019,Reina12021OVPML,Cichos2020,Dulaney2021,Colen2021} for the dynamics of soft matter and active matter by combining the variational method of approximation based on Onsager's variational principle (OVP)~\cite{Doi2015} with deep learning methods~\cite{ChunLiu2022}. Furthermore, the input of the neural networks should generally include both spatial and temporal coordinates; the output can be not only displacement fields, but also other slow variables such as polarization, concentration, \emph{etc}. In addition, we should also use and compare with the physics informed neural network (PINN)~\cite{Raissi2019} to solve the complicated Euler-Lagrange partial differential equations derived from variational principles.



\section*{Conflicts of interest}
There are no conflicts to declare. 

\section*{Acknowledgements}
X. Xu is supported in part by a project supported by the National Science Foundation for Young Scientists of China (NSFC, No.~12004082), by Guangdong Province Universities and Colleges Pearl River Scholar Funded Scheme (2019), by 2020 Li Ka Shing Foundation Cross-Disciplinary Research Grant (No.~2020LKSFG08A), by Provincial Science Foundation of Guangdong (2019A1515110809), by Guangdong Basic and Applied Basic Research Foundation (2020B1515310005), and by Featured Innovative Projects (No.~2018KTSCX282) and Youth Talent Innovative Platforms (No.~2018KQNCX318) in Universities in Guangdong Province.  
D. Wang acknowledges support from National Natural Science Foundation of China grant 12101524 and the University Development Fund from The Chinese University of Hong Kong, Shenzhen (UDF01001803).

\bibliography{refs} 

\end{document}